\documentclass[prd,twocolumn,superscriptaddress,floatfix,showpacs,preprintnumbers,amsmath,amssymb]{revtex4}
\usepackage[dvips]{graphicx}
\usepackage{epsfig}             % Allows inclusion of eps files.
\usepackage{dcolumn}            % used for alignment along decimal places in tables
\usepackage{bm}

\newcommand{\beq}[1]{\begin{equation} #1 \end{equation}}
\newcommand{\beqa}[1]{\begin{eqnarray} #1 \end{eqnarray}}
\newcommand{\greekbf}[1]{\bm{#1}}

\newcommand{\deriv}[2]{\frac{ d #1 }{ d #2 }}

\newcommand{\ibanez}{J. $\mathrm{M}^{\underline{a}}$ Ib\'{a}\~{n}ez}
\newcommand{\marti}{J. $\mathrm{M}^{\underline{a}}$ Mart\'{i}}

\newcommand{\aap}{{Astron.~Astrophys.}}                % Astronomy and Astrophysics
\newcommand{\physrep}{{Phys.~Rep.}}   % Physics Reports
\newcommand{\cmp}{{Comm.~Math.~Phys.}}  %Communications in Mathematical Physics
\newcommand{\cqg}{{Class.~Quant.~Grav.}}  %Classical and Quantum Gravity 
\newcommand{\grqc}[1]{{arxiv:gr-qc/#1}}
\newcommand{\lr}{{Living~Reviews}}

\newcommand{\pr}{{Phys.~Rev.}}                   % old Physical Review
\newcommand{\mcp}{{Meth.~Comp.~Phys.}}   %Methods of Computational Physics 

\begin{document}

\title{Driven neutron star collapse:\\ Type~I critical phenomena and the initial black hole mass distribution}

\date{\today}

\author{Scott C. Noble}
\email{scott-noble@utulsa.edu}
\homepage{http://personal.utulsa.edu/~scn759/}
\affiliation{
Department of Physics and Engineering Physics, 
The University of Tulsa,
800 South Tucker Drive,
Tulsa, OK 74104 USA}

\author{Matthew W. Choptuik}
\email{choptuik@physics.ubc.ca}
\homepage{http://laplace.physics.ubc.ca/People/matt/}
\affiliation{CIFAR Cosmology and Gravity Program, Department of Physics and Astronomy, 
University of British Columbia, 6224 Agricultural Road, Vancouver, Canada, V6T 1Z1}

%%%%%%%%%%%%%%%%%%%%%%%%%%%%%%%%%%%%%%%%%%%%%%%%%%%%%%%%%%%%%%%%%%%%%%%%%%%%%%%%%%%%%
%% ABSTRACT:   %%%%%%%%%%%%%%%%%%%%%%%%%%%%%%%%%%%%%%%%%%%%%%%%%%%%%%%%%%%%%%%
%%%%%%%%%%%%%%%%%%%%%%%%%%%%%%%%%%%%%%%%%%%%%%%%%%%%%%%%%%%%%%%%%%%%%%%%%%%%%%%%%%%%%

\begin{abstract}
We study the general relativistic collapse of neutron star (NS)
models in spherical symmetry.  Our initially stable models 
are driven to collapse by the addition of one of two things:
an initially ingoing velocity profile, or a shell of 
minimally coupled, massless scalar field that falls onto the star.  
Tolman-Oppenheimer-Volkoff (TOV) solutions with 
an initially isentropic, gamma-law equation of state serve
as our NS models.  The initial values of the 
velocity profile's amplitude and the 
star's central density span a parameter space which we have
surveyed extensively and which we find provides a rich picture of the 
possible end states of NS collapse.  This 
parameter space survey elucidates the boundary between 
Type~I and Type~II critical behavior in perfect fluids
which coincides, on the subcritical side, with the boundary 
between dispersed and bound end states.  
For our particular model, initial 
velocity amplitudes greater than $0.3c$ are needed to probe 
the regime where arbitrarily small black holes can form.
In addition, we investigate 
Type~I behavior in our system by varying the initial amplitude of 
the initially imploding scalar field.  In this case we find that 
the Type~I critical solutions resemble TOV solutions on the 
$1$-mode unstable branch of equilibrium solutions, and that 
the critical solutions' frequencies agree well with the 
fundamental mode frequencies of the unstable equilibria.  
Additionally, the critical solution's scaling exponent is shown to 
be well approximated by a linear function of the initial star's central density.  
\end{abstract}


\maketitle

%%%%%%%%%%%%%%%%%%%%%%%%%%%%%%%%%%%%%%%%%%%%%%%%%%%%%%%%%%%%%%%%%%%%%%%%%%%%%%%%%%%%%
%% CHAPTER:   %%%%%%%%%%%%%%%%%%%%%%%%%%%%%%%%%%%%%%%%%%%%%%%%%%%%%%%%%%%%%%%
%%%%%%%%%%%%%%%%%%%%%%%%%%%%%%%%%%%%%%%%%%%%%%%%%%%%%%%%%%%%%%%%%%%%%%%%%%%%%%%%%%%%%
\section{Introduction}
\label{sec:introduction}

The dynamics of compact gravitating objects out of equilibrium has always been a topic of much interest
in astrophysics.  Physical systems that fall under this subject
include supernovae, ``failed'' supernovae such as hypernovae or collapsars, gamma-ray 
burst (GRB) progenitors, coalescing binary neutron star
(NS) systems, accreting compact stars, and 
NSs that undergo sudden phase transitions, to name only a few.  
In the case of a core collapse supernova, a NS may form and undergo additional evolution. 
For instance, the outwardly-moving shock wave of matter from the supernova may stall and collapse onto the nascent neutron 
core \cite{zampieri-etal-1998}.  In contrast, if the NS  is in a binary system with a less 
compact companion star, accretion from the companion  may push the NS  over its 
Chandrasekhar limit.  In either of these cases, the resultant non-equilibrium system will most likely undergo  
a hydrodynamic implosion that will often result in black hole formation. 

Here we wish to present work that sets such excited NSs in the context of 
critical phenomena in general relativity.  Specifically, we wish to investigate 
1) the 
criteria required to initiate black hole formation, the boundary between black hole forming 
scenarios and those that do not form black holes, and 
2) the dynamical behavior of the systems in general.
This work is one of only several to date that ties critical phenomena to 
astrophysical scenarios
\cite{niemeyer-jedamzik-1998,niemeyer-jedamzik-1999,green-liddle,hawke-stewart-2002,novak,jin-suen-2006,wan-jin-suen-2008,liebling-lehner-neilsen-palenzuela-2010,radice-2010,kellerman-2010}.

We are certainly not the first to study numerical evolutions of NS models far from equilibrium.  
For example, Shapiro and Teukolsky \cite{shap-teuk-1980} asked whether a stable NS with a 
mass below the Chandrasekhar mass could
be driven to collapse by compression.  With a mixed Euler-Lagrangian code, they began to answer the question by studying 
stable stars whose density profiles had been ``inflated'' in a self-similar manner such that the 
stars became larger and more massive.  
Due to insufficient central pressure, such configurations were no longer 
equilibrium solutions and  inevitably collapsed.  By increasing the degree to which 
the equilibrium stars were inflated, they were able to supply more kinetic energy to the system. 
They found that black holes formed only for stars with masses greater than the 
maximum equilibrium mass.  In addition, Shapiro and Teukolsky studied accretion induced collapse, where it 
was again found that collapse to a black hole occurred only when the total mass of the system---in this case 
the mass of the star \emph{and} the mass of the accreting matter---was above the maximum stable mass.  
Both examples seemed to suggest that even driven
stars needed to have masses above the maximum stable mass in order to produce black holes.
Moreover, they only witnessed three types of outcomes: 1) homologous bounce, wherein
the entire star underwent a bounce after imploding to maximum compression; 2) non-homologous
bounce where less than $50\%$ of the matter followed a bounce sequence; and 3) direct collapse 
to a black hole.  
Also, Baumgarte et al.~\cite{baumgarte-etal-1995} using a Lagrangian code based on the 
formulation of Hernandez and Misner \cite{hernandez-misner} qualitatively confirmed these results. 

In order to investigate the question posed by Shapiro and Teukolsky further, Gourgoulhon 
\cite{gourg2,gourg1} used pseudo-spectral 
methods and realistic, tabulated equations of state to characterize the various ways in 
which a NS may collapse when given an \textit{ad hoc}, polynomial velocity profile. 
Such velocity profiles mimic those seen in core collapse simulations as described in 
\cite{may-white-paper,vanriper1}.  Given a sufficiently large amplitude of the profile, 
Gourgoulhon was able to form black holes from stable stars with masses well below the maximum.  
He was also able to observe bounces off the inner core, but was unable to continue the evolution
significantly past the formation of the shock since spectral techniques typically behave poorly for 
discontinuous solutions.  

To further explore this problem and resolve the shocks more accurately, Novak \cite{novak} used a 
Eulerian code with High-Resolution Shock-Capturing (HRSC) methods.  In addition, he surveyed the 
parameter space in the black hole-forming regime in much greater detail than previous studies, 
illuminating  a new scenario in which a black hole may form on the same dynamical time-scale as 
the bounce.  Depending on the 
amplitude of the velocity perturbation, such cases can lead to black holes that have smaller 
masses than their progenitor stars.  This dependence suggested that masses of black holes generated 
by NS collapse might not be constrained by the masses of their parents and, consequently, 
could---in principle---allow the black hole mass, $M_{\rm BH}$, to take on a continuum of values. 
In addition, in accordance with the study described in \cite{gourg2}, 
Novak found that the initial star did not have to 
be more massive than the maximum mass in order to 
evolve to a black hole.
In fact, he found that for two equations of state---the typical polytropic equation of state (EOS) and 
a realistic EOS described in \cite{pons-etal-2000}---arbitrarily small black holes could be made by 
tuning the initial amplitude of 
the velocity profile about the value at which black holes are first seen.  Hence, Novak's work
suggests that black holes born from NSs are able to have masses in the range 
$0 < M_{\rm BH} \le M_\star$, where $M_\star$ is the mass of the progenitor star.  This suggests 
that critical phenomena may play a role in the black hole mass function of driven NSs.

%TYPE-II
Critical phenomena in general relativity involves the study of the solutions---called \emph{critical} 
solutions---that lie at the boundary between 
black hole-forming and black hole-lacking spacetimes (for reviews please see 
\cite{choptuik-1998,gundlach,gundlach-rev2}).  
General relativistic critical phenomena  
began with a detailed numerical investigation of the dynamics of a minimally
coupled, massless scalar field in spherical symmetry~\cite{choptuik-1993}.
This first study identified three fundamental features of 
the critical behavior: 1) universality and 2) scale invariance 
of a critical solution that arises at threshold, with 
3) power-law scaling behavior in the vicinity of threshold.   
All three of these have now been seen in a multitude of collapse models with
a wide variety of matter sources, including 
perfect fluids~\cite{evans-coleman,neilsen-crit,brady_etal}, an~$\mathrm{SU}(2)$ 
Yang-Mills model~\cite{choptuik-chmaj-bizon,choptuik-hirshmann-marsa}, 
and collisionless matter \cite{rein-etal-1998,olabarrieta-choptuik} to cite
just a few.  It was eventually found that there are 
two related yet distinct types of critical phenomena, dubbed Type~I and Type~II, 
due to similarities 
between the critical phenomena observed in gravitational collapse, and 
those familiar from statistical mechanics.  

Type~II behavior entails critical solutions that are 
either continuously self-similar (CSS) or discretely self-similar (DSS).  
Supercritical solutions---those that form black holes---give rise to black holes with 
masses, $M_{\rm BH}(p)$,  that scale as a power-law, 
\beq{
M_\mathrm{BH}(p) \propto \left|p - p^\star\right|^\gamma  \quad ,
\label{mass-scaling}
}
implying that arbitrarily small black holes can be 
formed.  Here, $p$ parameterizes a $1$-parameter family of initial data with which one can tune toward 
the critical solution, 
located at $p=p^\star$, and $\gamma$ is the scaling exponent of the critical behavior.  
Since $M_\mathrm{BH}(p)$ is, loosely speaking, continuous across
$p=p^\star$, this type of critical behavior was named ``Type~II'' since it parallels 
Type~II (continuous) phase transitions in statistical mechanics.  

As in the statistical mechanical case, 
there is a Type~I behavior, where the black hole mass ``turns on'' at a finite value.
Type~I critical solutions are quite different from their Type~II counterparts, tending to be 
metastable star-like solutions that are either static or periodic.  
The critical solutions 
can therefore be described by a continuous or discrete symmetry in time, analogous to the Type~II
CSS and DSS solutions.  Unlike the Type~II case, however, the black hole masses of supercritical 
solutions do not follow a power-law scaling.  Instead, the span of time, 
$\Delta T_0(p)$---as measured by an 
observer at the origin---that a given solution is close to the critical solution scales with the 
solution's deviation in parameter space from criticality
\beq{
\Delta T_0(p) \propto - \sigma \ln \left| p - p^\star \right|  \quad , \label{type-i-scaling}
}
where $\sigma$ is the scaling exponent of Type~I behavior. 

We note that many of the features of critical gravitational collapse can be understood in a manner 
that also has a clear parallel in statistical mechanical critical phenomena.  In particular, 
the critical solutions that have been identified to date, although unstable, tend to be minimally 
so in the sense that they have only a single unstable mode in 
perturbation theory~\cite{evans-coleman,koike-et-al-1995}.  The Lyapunov exponent associated with 
this mode can then be directly related to the lifetime-scaling exponent, $\sigma$, for Type I 
solutions,
and to the mass-scaling exponent, $\gamma$, for Type II solutions.

%TYPE-I
In collapse models that involve matter characterized by one or more intrinsic length scales, the 
possibility of both types of critical behavior arises.
Indeed, the boundary separating the two types has been studied extensively in 
the SU(2) Einstein-Yang-Mills model \cite{choptuik-chmaj-bizon,choptuik-hirshmann-marsa} 
as well as the Einstein-Klein-Gordon system \cite{brady-chambers-goncalves}.  
In the latter case it was found that when 
the length scale $\lambda$---which characterized the ``spatial extent'' of 
a 2-parameter family of initial data---was small compared to the scale set by the massive scalar field,
Type~II behavior was observed.  The transition from Type~II to Type~I behavior was calculated for 
different families and was found to occur when $\lambda m \approx 1$, where $m$ is the 
(particle) mass of the scalar field. 

Two studies particularly close in spirit to our current work are due to 
Hawley and Choptuik \cite{hawley-choptuik-2000} and Lai and Choptuik~\cite{lai-choptuik}.
Instead of perturbing TOV solutions \citep{oppenheimer-volkoff,tolman-book,tolman-paper}, 
these authors perturbed stable, spherically symmetric, boson stars. 
Boson stars are self-gravitating configurations of a complex scalar field with some prescribed 
self-interaction (possibly just a mass term), whose 
only time-dependence is a phase that varies linearly with time (see \cite{jetzer-1992} 
and \cite{lee-pang-1992} for reviews).
For a given self-interaction, one can generically construct 
one-parameter families of boson stars, where the family parameter can conveniently be taken 
to be the central modulus, $\phi(0)$, of the complex field, and which plays the role of the central 
rest-mass density in TOV solutions.  As with their hydrostatic 
counterparts (discussed in more detail in Sec.~\ref{sec:init-star-solut}), when the total mass,
$M_\star(\phi(0))$, of the configurations is plotted as a function of $\phi(0)$, one typically 
finds a maximum mass for some $\phi(0)=\phi_{\rm max}(0)$ which signals a change in 
dynamical stability:
stars with $\phi(0)< \phi_{\rm max}(0)$ are stable, while those with $\phi(0) > \phi_{\rm max}(0)$
are unstable.  Additionally, for any family of boson stars, there is generally a branch of 
unstable stars---with $\phi(0)$ ranging from $\phi_{\rm max}$ to the next value where the mass function 
is a local minimum---that have precisely one unstable mode in perturbation theory.  These stars 
are thus candidates to be Type I critical solutions in a collapse scenario. 

Hawley and Choptuik 
perturbed  a boson star
by collapsing a spherical pulse of massless scalar field onto it from a distance sufficient to 
ensure that 
the two matter distributions were initially non-overlapping. 
As such a pulse collapses through the origin, the energy distributions associated with 
the two matter fields interact 
solely through the gravitational field.  For sufficiently large amplitudes 
of the scalar field, the resulting increase in curvature within the star is
enough to significantly compress it, ultimately resulting in either black hole formation
or a star that executes a sequence of oscillations, often of large amplitude.  
By tuning the initial amplitude 
of the scalar field, Type I critical solutions were found and, per the above observation,
were identified as (perturbed) one-mode unstable boson star configurations.  It was verified that the 
lifetimes of near-critical evolutions scaled according to~(\ref{type-i-scaling}), and 
that in each case the scaling exponent, $\sigma$, was consistent  with 
the inverse of the real part of the Lyapunov exponent, $\omega_{Ly}$, 
of the critical solution.  Furthermore, values of $\omega_{Ly}$ were independently calculated for several 
cases 
by applying linear perturbation theory to the static 
boson star backgrounds, and were shown to be in good agreement with those measured from the 
fully dynamical calculations.  Finally, since 
boson stars model many of the characteristics of TOV solutions, it was 
conjectured that the observed critical behavior would carry over to the fluid case.  

We note that in the results reported in~\cite{hawley-choptuik-2000} the end state of 
marginally subcritical collapse was {\em not} identified as a periodic spacetime (i.e.~a perturbed 
boson star);  rather, it was assumed that the stars 
would disperse to spatial infinity in such cases.  Upon evolving subcritical configurations for 
longer physical times, 
Lai and Choptuik \cite{lai-choptuik}---in work performed simultaneously to that of \cite{noble}---found 
that the end states were, in fact, gravitationally bound and oscillatory.  
These results were subsequently verified by  Hawley \cite{hawley-pc}.
Interestingly, in both studies it was found that during the non-trivial gravitational interaction 
of the massless scalar field 
and the boson star there was a transfer of mass-energy from the massless scalar field to the complex 
scalar field, resulting in an increase of
the gravitating mass of the boson star.

Returning now to the fluid case, 
Siebel et al.~\cite{siebel-font-pap-2001} sought to measure the maximum NS mass allowed by  the 
presence of a perturbing pulse of minimally-coupled, massless scalar field.  A general relativistic hydrodynamic 
code using a characteristic formulation was used to investigate the spherically symmetric system.
However, instead of varying the massless scalar field
they studied five distinct star solutions having a range of central densities that straddled 
the threshold of black hole formation.  
They found that the perturbation either led to a black hole or to oscillations of the star about its initial 
configuration.   Further, in order to test their new 3-dimensional general relativistic fluid 
code, Font et al.~\cite{font-etal2} dynamically calculated the fundamental and harmonic mode frequencies
of spherical TOV solutions. 
They observed the transition of a TOV solution on the \emph{unstable} branch to the 
stable branch by evolving an unstable solution that was perturbed at the truncation error level.  
The unstable star overshot and then oscillated about the stable solution, contradicting a common 
assumption in the field that stars from the unstable branch always formed black holes.  Evolving from initial conditions 
consisting of an unstable TOV star
has continued to be used for code-testing purposes~\cite{radice-rezzolla-2011}.
In \cite{liebling-lehner-neilsen-palenzuela-2010} Liebling et al. performed a similar study 
with weakly magnetized unstable TOV solutions in 3-d, but 
employed explicit and tunable perturbations to the pressure and density.  They, too, found evidence for Type~I behavior, though 
were unable to tune sufficiently close to the threshold to demonstrate the expected scaling behavior.  All the different kinds of 
perturbations they employed drove the system to the same, seemingly \emph{universal} solution.  Proximity to the critical 
threshold was improved  in \cite{radice-2010}, wherein they perturbed axisymmetric unstable TOV stars by truncation error 
and a small ingoing velocity distribution, while tuning with the central density of the star. 

Apart from the work presented here (and here~\cite{noble}), the most exhaustive 
explorations of Type~I behavior involving NSs are that of Jin and Suen~\cite{jin-suen-2006},  
Wan et al.~\cite{wan-jin-suen-2008,wan-submitted-2010,wan-thesis-2010}, and Kellerman, Radice, and Rezzolla~\cite{kellerman-2010}.  
The results presented in these papers indicated that the head-on 
axisymmetric collision of two NSs can be tuned with a variety of initial data parameters to a critical threshold that bifurcates end states 
involving either a single black hole or a single more massive oscillating NS.  
Universality of the critical behavior was supported by 
tuning separately the initial magnitude of the stellar velocities, central densities and adiabatic index of their polytropic EOS.  
Threshold solutions were found to high precision for all three of the tuning variables.  All threshold solutions were found 
to be perturbed TOV solutions on the unstable branch, no matter the tuning parameter. 
 Since changes in the adiabatic index
may mimic the effects of cooling and accretion, an interesting conjecture 
was made that critical behavior might be realizable without 
the need for fine tuning \cite{jin-suen-2006}.  
Further, frequencies at which the near-threshold solution 
oscillated were measured and found to differ---by one to two orders of magnitude---from the frequencies of the $l=0,1$ perturbation modes 
about the \emph{initial stable} TOV solution \cite{wan-jin-suen-2008}.  The seeming discrepancy in frequencies was eventually 
explained by \cite{kellerman-2010} when they demonstrated that the near-threshold solutions were perturbed TOV solutions on the 
unstable branch, and that the oscillations occurred at the fundamental mode of the unstable TOV solution---not the original stable TOV 
solution.  This realization in the literature paralleled conclusions made years before in the boson star context \citep{lai-choptuik}, and 
in the TOV context \citep{noble}. 

In this work, we investigate both types of critical behavior using a perfect fluid model,
although we focus for the most part on the Type~I case.  For the first time with TOV solutions, we demonstrate that 
the scaling exponent, $\sigma$, is consistent  with 
the inverse of the real part of the Lyapunov exponent, $\omega_{Ly}$, 
of the critical solution.  This provides further evidence to support the notion that the Type~I critical solutions 
are perturbed TOV solutions on the unstable branch. 
The initial conditions which we adjust entail a stable TOV star with the stiffest causal polytropic EOS ($\Gamma = 2$), 
plus some sort of ``perturbing agent.''
The methods by which we drive a star to a non-equilibrium 
state involve: 1) giving the star an initially ingoing velocity profile, and 2) collapsing 
a spherical shell of scalar field onto it.  Neither method can be considered truly perturbative 
since both can drive the star to total obliteration or prompt collapse to a black hole, but 
we use this term since a better one is lacking. 

Sec.~\ref{sec:theoretical-model} provides  the 
theory describing our systems and the numerical methods we use to simulate them.
In Sec.~\ref{sec:veloc-induc-neutr},
we begin our study of stellar collapse by extensively covering the parameter space of initial 
conditions for velocity-perturbed stars.  The results from this section provide a broad view of the 
range of dynamical scenarios one can expect in the catastrophic collapse of NSs.  We then 
employ this knowledge in our examination of the solutions 
that lie on the verge of black hole formation.  Both Type~I 
and Type~II solutions are found.  The stars' Type~I critical behavior 
is explored in Sec.~\ref{sec:type-i-critical} (their  Type~II behavior 
has been investigated in a related paper \cite{noble-choptuik1}). 
The threshold solutions we calculate from the Type~I study are then compared to 
unstable TOV solutions.  In addition, for the first time, a parameter-space boundary 
separating the two types of phenomena is identified and discussed. 
Finally, we conclude in Sec.~\ref{sec:concl-future-work} with some closing remarks and 
notes on possible future work. 

%%%%%%%%%%%%%%%%%%%%%%%%%%%%%%%%%%%%%%%%%%%%%%%%%%%%%%%%%%%%%%%%%%%%%%%%%%%%%%%%%%%%%
%%%%%%%%%%%%%%%%%%%%%%%%%%%%%%%%%%%%%%%%%%%%%%%%%%%%%%%%%%%%%%%%%%%%%%%%%%%%%%%%%%%%%
%%%%%%%%%%%%%%%%%%%%%%%%%%%%%%%%%%%%%%%%%%%%%%%%%%%%%%%%%%%%%%%%%%%%%%%%%%%%%%%%%%%%%
%% Section:   %%%%%%%%%%%%%%%%%%%%%%%%%%%%%%%%%%%%%%%%%%%%%%%%%%%%%%%%%%%%%%%
%%%%%%%%%%%%%%%%%%%%%%%%%%%%%%%%%%%%%%%%%%%%%%%%%%%%%%%%%%%%%%%%%%%%%%%%%%%%%%%%%%%%%
\section{Theoretical Model}
\label{sec:theoretical-model}

The equations and methods employed in this study closely follow those used in \cite{noble-choptuik1}. 
The primary difference is that we sometimes use a massless scalar field that is minimally coupled 
to gravity, and hence to the fluid.  We refer the reader to \cite{noble-choptuik1} 
for details regarding the evolution of the hydrodynamics equations, but 
give here the equations that describe this ``fluid+scalar'' system and the methods used to evolve 
the scalar field.

%%%%%%%%%%%%%%%%%%%%%%%%%%%%%%%%%%%%%%%%%%%%%%%%%%%%%%%%%%%%%%%%%%%%%%%%%%%%%%%%%%%%%
%%%%%%%%%%%%%%%%%%%%%%%%%%%%%%%%%%%%%%%%%%%%%%%%%%%%%%%%%%%%%%%%%%%%%%%%%%%%%%%%%%%%%
\subsection{The Geometry Equations}
\label{sec:geometry}

We largely follow the notation established in our previous paper on Type II collapse 
of a perfect fluid~\cite{noble-choptuik1}.  We use geometrized units such that $G = c = 1$,  and 
tensor notation and sign conventions that follow  Wald~\cite{wald}.  When coordinate bases are explicitly used, Greek and Roman indices
will refer to spacetime and purely spatial components, respectively 
(i.e. $\mu, \nu, \ldots \in \{0, 1, 2, 3\}$, and
$i, j, k \in \{1, 2, 3\}$).  
Quantities in bold-face, e.g. $\mathbf{q}, \mathbf{f}$, are generally state vectors. 

As in many previous critical phenomena studies in spherical symmetry  
\cite{brady_etal,choptuik-1993,choptuik-chmaj-bizon,neilsen-crit,novak}, we employ the so-called 
polar-areal metric
\beq{ 
ds^2 = - \alpha\left(r,t\right)^2 dt^2 + a\left(r,t\right)^2 dr^2 
+ r^2 d\Omega^2  \quad . \label{metric}
}
Since we will use a variety of sources in this study, we state the equations governing 
the metric functions using the formulation of Arnowitt, Deser and Misner (ADM) \cite{adm} and 
no specific assumption about the 
precise form of stress-energy tensor.  To update $a$ at each time step, we solve the 
Hamiltonian constraint,
\beq{
\frac{a^\prime}{a} = 4 \pi r a^2 \varrho + \frac{1}{2r}\left(1 - a^2\right)\quad , 
\label{polar-areal-hamiltonian-const}
}
where $\varrho$ is the local energy density measured by an observer moving orthonormal to the
spacelike hypersurfaces.  Note that a ``prime'' will denote differentiation with respect to 
$r$ and a ``dot'' will represent differentiation with respect to $t$. 
In our coordinate basis, the 4-velocity, $n^a$, of this orthonormal observer has components
\beq{
n^\mu = \left[ \frac{1}{\alpha}, 0 , 0 , 0 \right]^T \quad . \label{normal-4-velocity}
}
Hence, $\varrho$ can be shown to be
\beq{
\varrho = T_{\mu \nu} n^\mu n^\nu = T_{t t} / \alpha^2 . \label{adm-density-eval}
}
The lapse function $\alpha$ is calculated at each step via the polar slicing condition,
\beq{
\frac{\alpha^\prime}{\alpha} = \frac{a^\prime}{a}  + \frac{1}{r} \left(a^2 - 1 \right) 
- \frac{8 \pi a^2}{r} \left[ T_{\theta \theta} - \frac{r^2}{2} \left( {T^i}_i - \varrho \right) \right] 
\quad . \label{polar-areal-slicing-condition}
}
Even though it is  used solely for diagnostic purposes, we state here for completeness the momentum constraint, 
which yields an evolution equation for $a$,
\beq{
\dot{a} = - 4 \pi r \alpha a j_r  \quad , \label{polar-areal-momentum-const}
}
where  $j_r$ is the only non-vanishing component of the momentum density measured by the orthonormal observer, 
\beq{
j_a \equiv \left( g_{a c} + n_a n_c \right) n_b T^{b c} \quad . \label{adm-momentum-def}
}
For diagnostic purposes, it is convenient to introduce the
mass aspect function,  $m$, given by
\beq{  
m(r,t) \equiv \frac{r}{2} \left( 1 - \frac{1}{a^2} \right) \, . \label{massaspect}
}
We note that polar-areal coordinates {\em cannot} penetrate apparent horizons, but 
that the formation of a black hole in a given calculation is nonetheless 
signaled by $2m(t,{\tilde r})/{\tilde r}\rightarrow~1$, for some specific
radial coordinate, $r={\tilde r}$.

%%%%%%%%%%%%%%%%%%%%%%%%%%%%%%%%%%%%%%%%%%%%%%%%%%%%%%%%%%%%%%%%%%%%%%%%%%%%%%%%%%%%%
%%%%%%%%%%%%%%%%%%%%%%%%%%%%%%%%%%%%%%%%%%%%%%%%%%%%%%%%%%%%%%%%%%%%%%%%%%%%%%%%%%%%%
\subsection{The Matter Equations}
\label{sec:matter-equations}

We model NS matter as a perfect fluid. 
Modern conservative methods that utilize the characteristic structure of 
the fluid equations of motion expressed in {\em conservative} form have been
very successful in evolving
highly-relativistic flows in the presence of strong 
gravitational fields (see  \cite{banyuls,font-review,font-etal2,neilsen,romero} 
for a small but representative selection of papers on this topic), and 
we follow that approach here.
In particular, we use a formulation used by
Romero et al.~\cite{romero} and a change of variables similar to that
performed by Neilsen and Choptuik \cite{neilsen}. 

One way in which we drive NS models to collapse entails the inclusion of a massless 
scalar field which dynamically perturbs the star.  We also use a driving mechanism
that involves no scalar field.
Not surprisingly, it turns 
out that the equations governing the geometry and fluid equations in the ``fluid-only'' system can be recovered
from those in the ``fluid+scalar'' system simply by setting the scalar field, 
$\phi(r,t)$, to zero for all $r$ and $t$.
Hence, our numerical implementation always uses the full ``fluid+scalar'' equations for determining fluid and geometric
fields: if we wish to include the scalar field, we simply initialize it to a non-zero value and evolve 
it in tandem with the fluid.  Thus, by stating the fluid equations of motion
(EOM) for the ``fluid+scalar'' system, we are also 
simultaneously---yet indirectly---stating them in the ``fluid-only'' system.

The EOM for the two matter sources are derived, in part, from the local conservation of energy 
\beq{
\nabla_a {T^a}_b = 0  \quad , \label{energycons}
}
where $T_{a b}$ is the \emph{total} stress-energy tensor.  Since there is no explicit coupling between the 
two matter sources, the total stress tensor is a sum of the stress tensors of the individual sources
\beq{
T_{a b} = {\tilde{T}}_{a b} + {\hat{T}}_{a b}  \quad , \label{totalstress}
}
where $\hat{T}_{a b}$ and $\tilde{T}_{a b}$  are the stress-energy tensors of the fluid and scalar field, respectively.
Further, the local conservation of energy equation holds \textit{separately} for each stress-energy. 
Specifically,
\beq{
\nabla^a T_{a b} = \nabla^a {\tilde{T}}_{a b} = \nabla^a {\hat{T}}_{a b} = 0 \ .
\label{bothenergyconservation}
}

The scalar field stress-energy tensor is
\beq{
{\tilde{T}}_{a b} = \nabla_a \phi \nabla_b \phi - \frac{1}{2} g_{a b} 
\left(\, \nabla_c \phi \nabla^c \phi + 2 V(\phi) \, \right)   \quad , 
\label{scalarstress}
}
where $V(\phi)$ is the scalar potential. In the following equations, 
we will assume that $V(\phi)$ is non-zero, however, we have set $V(\phi) \equiv 0$ in all 
of the  calculations reported below.
Since there is no direct interaction between the scalar field and the fluid, 
(\ref{bothenergyconservation}) yields the usual 
equation of motion for the scalar field:
\beq{
\Box \phi \equiv \nabla^a \nabla_a \phi = \partial_{\phi} V(\phi)  \quad . \label{generalscalareom} 
}
We can convert this to a system of first-order (in time) PDEs by introducing auxiliary 
variables, $\Xi$ and $\Upsilon$, defined by
\beq{ 
\Xi \equiv \phi '  \quad , \quad \Upsilon \equiv \frac{a}{\alpha} \dot{\phi} \quad . 
\label{Xi-Upsilon}
}
With these definitions the EOM become
\beq{
\dot{\Xi} = \left( X \Upsilon \right) '  \quad , \label{scalareom1}
}
\beq{
\dot{\Upsilon} = \frac{1}{r^2} 
\left( r^2 X \Xi \right) ' - \alpha a \partial_\phi V  \quad , 
\label{scalareom2}
}
where $X \equiv \alpha / a$. 

The fluid equations of motion can be easily derived from the definition of the 
perfect fluid stress-energy tensor,
\beq{
\hat{T}_{a b} = \left( \rho + P \right) u_a u_b + P g_{a b} \quad , \label{stress}
}
the local conservation of energy equation (\ref{bothenergyconservation})
and the \emph{local conservation of baryon number}
\beq{
\nabla_a  \left( \rho_\circ u^a \right)  = 0  \quad . \label{currentcons}
}
Here, $u^a$ is the 4-velocity of a given fluid element, $P$ is the isotropic pressure, 
$\rho = \rho_\circ \left(1 + \epsilon\right)$ is the
energy density, $\rho_\circ$ is the rest-mass energy density, and $\epsilon$ is the specific internal energy.
Instead of the 
$4$-velocity of the fluid, a more useful quantity is the radial component of the Eulerian velocity 
of the fluid as measured by a Eulerian observer:
\beq{
v = \frac{ a u^r }{\alpha u^t}  \quad , 
}
where $u^\mu = \left[ u^t, u^r, 0, 0\right]$ (recall that we are working in spherical symmetry).  
The associated ``Lorentz gamma function'' is defined by 
\beq{ 
W = \alpha u^t  \quad . 
\label{w}
}
Given the fact that the 4-velocity is time-like and unit-normalized, i.e. $u^\mu u_\mu=-1$, 
$v$ and $W$ are related by
\beq{
W^2 = \frac{1}{1 - v^2} \quad . \label{vwrelation}
}

In conservation form, the fluid's EOM are 
\beq{ 
\partial_t \mathbf{q} 
+ \frac{1}{r^2} \partial_r \left( r^2 X \mathbf{f} \right) = \greekbf{\psi} 
\quad , \label{conservationeq}
}
where the state vector $\mathbf{q}$ is a vector of \emph{conserved} variables,  and 
$\mathbf{f}$ and $\greekbf{\psi}$ are---respectively---the flux and source state vectors. 
Our choice of conserved variables follows that of Neilsen and Choptuik \cite{neilsen}, and leads to 
improved accuracy in the ultrarelativistic regime ($\rho \gg \rho_0$): 
\beq{ 
\mathbf{q} =  \left[ \begin{array}{c} D \\ \Pi \\ \Phi \end{array}\right]  
 \ , \ 
\mathbf{f} = \left[ \begin{array}{c} D v \\ v \left( \Pi + P \right) + P
        \\ v \left( \Phi + P \right) - P \end{array} \right]
\ , \ 
\greekbf{\psi} = \left[ \begin{array}{c} 0 \\ \Sigma \\ -\Sigma \end{array} \right] 
\quad , 
\label{ideal-piphi-state-vectors}
}
where 
\begin{eqnarray}
D & = & a \rho_{\circ} W \label{D}  \quad , \\ 
\Pi  & = &  E - D + S \quad , \label{Pi} \\
\Phi & = &  E - D - S \quad , \label{Phi} \\ 
S & = & \rho_\circ h W^2 v \label{S} \quad , \\
E & = & \rho_\circ h W^2 - P \label{E} \quad ,
\end{eqnarray}
$h \equiv 1 + \epsilon + P/\rho_\circ$ is the the specific enthalpy of the fluid, 
$D$ is the Eulerian rest-mass density, and $\Pi$ and $\Phi$ are linear combinations of the 
Eulerian momentum density ($S$) and internal energy density ($E-D$). 
We use $P$, $\rho_\circ$, and $v$ as primitive variables. 
For the sake of efficiency, we state the source function, $\Sigma$, 
in terms of derivatives of the metric functions so that additional 
matter sources can be incorporated into the model more easily:
\beq{ 
\Sigma \equiv \Theta + \frac{2 P X}{r} \label{Sigma}
}
and 
\beq{
\Theta = - \frac{2 \dot{a} S}{a} - \frac{ \alpha^\prime }{\alpha} X E - \frac{a^\prime}{a} X \left( S v + P \right) 
\quad.  \label{Theta-general}
}
In practice, we use a simplified form of $\Theta$ derived from the constraints 
(\ref{polar-areal-hamiltonian-const},\ref{polar-areal-momentum-const}) and the slicing condition 
(\ref{polar-areal-slicing-condition}) to eliminate $a^\prime$, $\alpha^\prime$ and $\dot{a}$.   However, this 
requires knowledge of the full stress-energy tensor, $T_{a b}$, not just the fluid's stress-energy
tensor, $\hat{T}_{a b}$, to calculate.  In the ``fluid+scalar'' system, 
\beqa{
\Theta & = & \alpha a \left\{  \left( S v - E \right) \left[ 
4 \pi r \left( 2 P - V(\phi) \right) + \frac{m}{r^2} \right] \right. \nonumber \\ 
& + & \left. P \left( \frac{m}{r^2} - 4 \pi r V(\phi) \right) \right\} \label{Theta-both} \\
& - & \, 2 \pi r X \left[ 4\, \Xi \Upsilon S \, + \, 
\left( \Xi^2 + \Upsilon^2 \right)
\left( S v + P + E \right) \right] \quad . \nonumber 
}

When following the gravitational interaction between the fluid and scalar field, particularly
interesting quantities to track are the two contributions to $dm/dr$:
\beq{
\deriv{ m }{ r } \ = \ 4 \pi r^2 \varrho \ = \  4 \pi r^2 \varrho_{\,_{\mathrm{fluid}}}
  + \ 4 \pi r^2 \varrho_{\,_{\mathrm{scalar}}}  \quad , \label{dmdr1}  
}
\beq{
\deriv{ m_{\mathrm{fluid}} }{ r }  =  4 \pi r^2 E \quad , \label{dmdr-fluid}
}
\beq{
\deriv{ m_{\mathrm{scalar}} }{ r }  = 
 4 \pi r^2 \left[ \frac{1}{2 a^2} \left( \Xi^2 + \Upsilon^2 \right) + V(\phi) \right] 
\quad  .  \label{dmdr-scalar}
}
However, the two mass contributions can only be unambiguously differentiated in regions 
of non-overlapping support, since---for instance---$\partial m_\mathrm{scalar} / \partial r$ depends on metric
quantities which in turn depend on the local energy content of all matter distributions that are present.
We note that expressing $\deriv{ m }{ r }$ in the form of Eq.~(\ref{dmdr1}) is possible because of our 
particular gauge choice. 

The EOS closes the 
system of hydrodynamic equations.  Because of the extensive nature of our parameter space survey, 
we wish to restrict ourselves to closed-form (i.e. non-tabulated) state equations.  For isentropic flows, 
the polytropic EOS, 
\beq{
P = K \rho_\circ^\Gamma  \quad , \label{polytrope-eos} 
}
for some constant, $K$, and adiabatic index, $\Gamma$,  is commonly used.  In addition, we use 
the ``ideal-gas'' or ``gamma-law'' EOS
\beq{
P=\left(\Gamma-1\right)\rho_\circ \epsilon \quad .
\label{ideal-eos}
}
Our initial NS models are solutions to the spherically-symmetric 
hydrostatic Einstein equations, and are commonly known as Tolman-Oppenheimer-Volkoff (TOV) solutions 
\cite{oppenheimer-volkoff,tolman-paper}.  We use both EOSs (\ref{ideal-eos},\ref{polytrope-eos}) to set 
the initial data, but use only the ideal-gas EOS (\ref{ideal-eos}) to evolve  
any specific configuration \footnote{A prescription for 
numerically solving the TOV equations can be found in \cite{shapiro-and-teuk}.}.  To simulate stiff
matter at super-nuclear densities---characteristic of neutron stars---we use $\Gamma=2$ in all 
of the calculations described below.
We also note that, as pointed out by Cook et al.~\cite{cook-shap-teuk-1992}, the constant $K$ can be 
thought of as the fundamental length scale of the system, with which one can use to scale any 
dynamical quantity with values of $(K,\Gamma)$ to a system with different values $(K',\Gamma')$.
As with $G$ and $c$, we set $K=1$. This makes our equations unitless, ensuring that our dynamical
variables are not at arbitrarily different orders of magnitude, and, as discussed in 
App.~\ref{app:unit-conversion}, expediting the transformation of results to another 
set of $(K,\Gamma)$.

In summary, 
in our simulations of  self-gravitating, ideal-gas fluids, the fluid is evolved by solving
(\ref{conservationeq},\ref{ideal-piphi-state-vectors}), the scalar field is evolved
using (\ref{scalareom1}-\ref{scalareom2}), while the geometry is simultaneously
calculated using the Hamiltonian constraint (\ref{polar-areal-hamiltonian-const}) and the slicing condition 
(\ref{polar-areal-slicing-condition}).  The specific methods we employ to numerically integrate these equations are 
briefly explained in Sec.~\ref{sec:numerical-techniques}.  

%%%%%%%%%%%%%%%%%%%%%%%%%%%%%%%%%%%%%%%%%%%%%%%%%%%%%%%%%%%%%%%%%%%%
\subsection{Initial Star Solutions}
\label{sec:init-star-solut}

Since the TOV equations take the form of a coupled set of ODEs, their solution does not generally
require the use of sophisticated numerical methods.  Readers who are interested in more details 
are referred to the pseudo-code description in Shapiro and Teukolsky~\cite{shapiro-and-teuk}, 
as well as the discussion of our specific approach given in~\cite{noble}.

\begin{figure}[htb]
\includegraphics[scale=0.4]{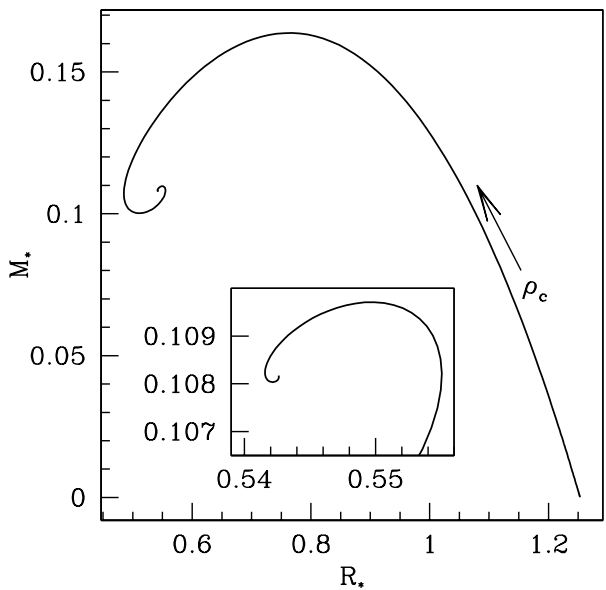}
\caption{Mass versus radius of TOV solutions using $\Gamma=2$ and $K=1$ with the 
polytropic EOS (\ref{polytrope-eos}). In the inset, we show a detailed view of the spiraling behavior. 
The arrow along the right side of the curve indicates the direction of increasing central density. 
\label{fig:tov-m-v-r}}
\end{figure}

Analysis of TOV solutions has a rich history \cite{harrison-etal} which we will not discuss here. 
We do, however, wish to 
note one important aspect of such solutions that is crucial to understanding their 
role in 
Type~I critical behavior, and which has already been touched upon in the Introduction in the
context of boson stars.  Given an EOS, the TOV solutions can be parameterized by their central 
pressures; in our case, the EOS (\ref{polytrope-eos}) allows us to reparameterize the solutions 
with respect to the central rest-mass density, $\rho_c$.  Arguments from linear stability analysis \cite{harrison-etal}
tell us that TOV solutions with the smallest central densities are stable to small perturbations, while those solutions
with $\rho_c$ at the opposite end of the spectrum (large $\rho_c$) are unstable.  
A plot of $M_\star(R_\star)$ is shown in Fig.~\ref{fig:tov-m-v-r}, where $R_\star$ is the radius of the star
and we see that $M_\star(R_\star)$ winds-up 
with increasing central density.  At the global maximum of $M_\star(R_\star)$ the fundamental, or lowest, mode 
becomes unstable.  After each subsequent local extremum in the direction of increasing
$\rho_c$, the next lowest mode becomes unstable.  For instance, there are four local extrema of
$M_\star(R_\star)$ shown in Fig.~\ref{fig:tov-m-v-r}, so those solutions with the largest $\rho_c$ will 
have their four lowest modes exponentially grow in time.  

As discussed previously, black hole critical solutions are typically characterized by a single growing mode.
Hence, the Type~I behavior associated with ``perturbed'' TOV solutions can be immediately anticipated to entail 
those TOV solutions that lie between the first and second extrema of $M_\star(\rho_c)$.  
For subsequent reference
we note that with the units and EOS that we have adopted, the most massive stable TOV solution has 
a central density $\rho_c \simeq 0.318$ and
a mass $M_\star \simeq 0.1637$.

After the initial, star-like solution is calculated, an ingoing velocity profile
is sometimes added to drive the star to collapse.  In order to do this, we follow
the prescription used in \cite{gourg2} and \cite{novak}.  The method described therein involves specifying the 
coordinate velocity, 
\beq{
  U \equiv \frac{ d r }{ d t } = \frac{ u^r }{ u^t }   \quad ,
  \label{radial-coord-velocity}
}
of the star, and then finding the Eulerian velocity, $v$, once the geometry has been calculated.  
In general, the profile takes the algebraic form
\beq{ 
U(x) = A_0 \left( x^3 - B_0 x \right) \quad . \label{v-profile-general}
} 
The two profiles that were used in \cite{novak} are 
\beqa{
U_1(x) & = &  \frac{U_\circ}{2} \left( x^3 - 3 x \right) \quad , \nonumber \\ 
U_2(x) & = &  \frac{27 \, U_\circ}{10 \sqrt{5}} \left( x^3 - \frac{5 x}{3} \right)  
\quad , \label{v-profile-12} 
} 
where $x \equiv r / R_\star$. Unless stated otherwise, $U_1$  will be used for all the results herein. 

Specifying the coordinate velocity instead of $v$ 
complicates the computation of the metric functions at $t=0$.  
Our method for dealing with this difficulty
is described in App.~\ref{sec:calc-init-star}. 

%%%%%%%%%%%%%%%%%%%%%%%%%%%%%%%%%%%%%%%%%%%%%%%%%%%%%%%%%%%%%%%%%%%%%%%%%%%%%%%%%%%%%
%%%%%%%%%%%%%%%%%%%%%%%%%%%%%%%%%%%%%%%%%%%%%%%%%%%%%%%%%%%%%%%%%%%%%%%%%%%%%%%%%%%%%
%%%%%%%%%%%%%%%%%%%%%%%%%%%%%%%%%%%%%%%%%%%%%%%%%%%%%%%%%%%%%%%%%%%%%%%%%%%%%%%%%%%%%
%% CHAPTER:   %%%%%%%%%%%%%%%%%%%%%%%%%%%%%%%%%%%%%%%%%%%%%%%%%%%%%%%%%%%%%%%
%%%%%%%%%%%%%%%%%%%%%%%%%%%%%%%%%%%%%%%%%%%%%%%%%%%%%%%%%%%%%%%%%%%%%%%%%%%%%%%%%%%%%
\section{Numerical Techniques}
\label{sec:numerical-techniques}

Simulating the highly-relativistic flows encountered in the driven
collapse of NSs entails solving a system of coupled, partial and ordinary 
differential equations that describe how the fluid, scalar field, and gravitational field evolve in time.  
High-resolution shock-capturing methods are used to evolve the fluid and the Iterative Crank Nicholson
method, with second-order spatial differences, is used for the scalar field.  
Both methods are second-order accurate, except that the 
fluid method is first-order accurate near shocks and at local extrema.  The Rapid Numerical Prototyping Language (RNPL) 
written by Marsa and Choptuik \cite{marsa-choptuik} is used 
to handle check-pointing, input/output, and memory management for all our simulations; we do not use the language
of RNPL itself for our finite differencing, but use original, secondary routines that are called 
from the primary RNPL routines. More details of the code, along with descriptions of code tests 
can be found in \cite{noble,noble-choptuik1}.

%%%%%%%%%%%%%%%%%%%%%%%%%%%%%%%%%%%%%%%%%%%%%%%%%%%%%%%%%%%%%%%%%%%%
%%%%%%%%%%%%%%%%%%%%%%%%%%%%%%%%%%%%%%%%%%%%%%%%%%%%%%%%%%%%%%%%%%%%
%%%%%%%%%%%%%%%%%%%%%%%%%%%%%%%%%%%%%%%%%%%%%%%%%%%%%%%%%%%%%%%%%%%%
\section{Velocity-induced Neutron Star Collapse}
\label{sec:veloc-induc-neutr}

Here we present a description of the various dynamic
scenarios we have seen in perturbed NS models, 
as a function of the initial star solution and the magnitude of the initial 
velocity profile.  These results are compared to those from previous studies---most notably
that of Novak \cite{novak}---but also provide some new insights.
Specifically, this section provides a description of various phases we have identified in parameter space, 
including those from a survey of the subcritical regime that is more detailed than has
been reported in prior work. 
In Sec.~\ref{sec:type-i-critical} we then focus on the critical phenomena observed at 
the threshold of black hole formation, and where collapse is induced via interaction of the 
fluid star with a collapsing pulse of massless scalar field.

In this section any specific TOV solution is driven out of equilibrium
by endowing it with an ingoing 
profile for the initial coordinate velocity, $U(r,0)$, as described in Sec.~\ref{sec:init-star-solut}.
We measure the magnitude of this perturbation by the absolute value of the minimum value of the 
Eulerian velocity $v$, $v_{\min}$,  at the initial time.  We find that $v_{\min}$ is 
uniquely specified by the parameter $U_\circ$ provided that we
follow the prescription for generating perturbed TOV stars 
given in App.~\ref{sec:calc-init-star}.  We also note that  $v_{\min}$ is a more physical
quantity than similar parameters---e.g. $U_\circ$---that pertain to the fluid's
gauge-dependent, coordinate velocity.  

\begin{figure}[htb]
\includegraphics[scale=0.4]{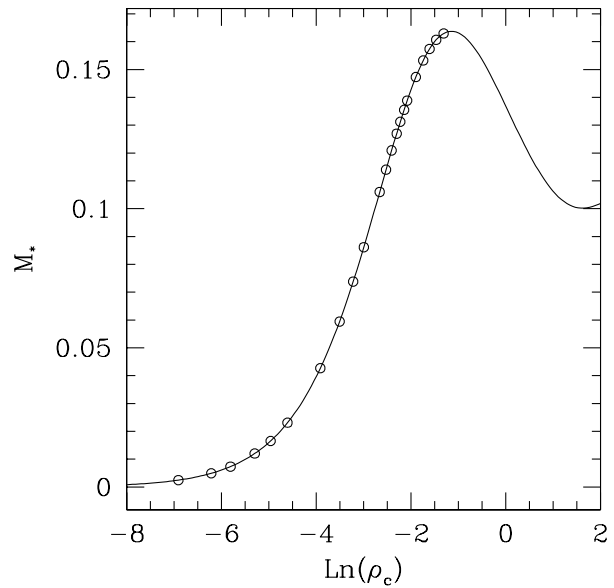}
\caption{Initial TOV solutions used in the parameter space survey.
\label{fig:rho-params}}
\end{figure}

Our survey used 22 different stable TOV solutions---specified by the initial central density $\rho_c$---shown in 
Fig.~\ref{fig:rho-params}.  The solutions used for the parameter space 
survey are displayed along the $M_\star(\rho_c)$ curve for $\Gamma=2$ TOV solutions.  We note 
that a wide spectrum of  stars were chosen, from non-compact stars that are relatively 
large and diffuse, to compact and dense stars.  

\begin{figure}[htb]
\includegraphics[scale=0.4]{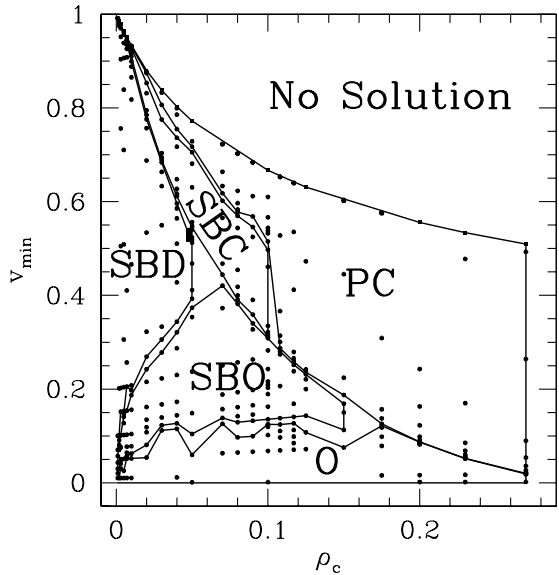}
\caption{Parameter 
space showing the regions in which various outcomes (phases) occur. The space is spanned by 
the initial magnitude of the velocity perturbation, $v_{\min}$, and 
the initial central density of the star, $\rho_c$. 
The small black rectangular region located at $(\rho_c,v_{\min})\sim (0.05,0.53-0.55)$ represents a set of solutions that 
undergo an SBO-type evolution.  Phase Legend: PC = Prompt Collapse, 
SBC = Shock-Bounce-Collapse, SBD = Shock-Bounce-Dispersal,
SBO = Shock-Bounce-Oscillation, O = Oscillation.  See text for further 
explanation of the various phases.
\label{fig:pspace}}
\end{figure}

\begin{figure}[htb]
\includegraphics[scale=0.4]{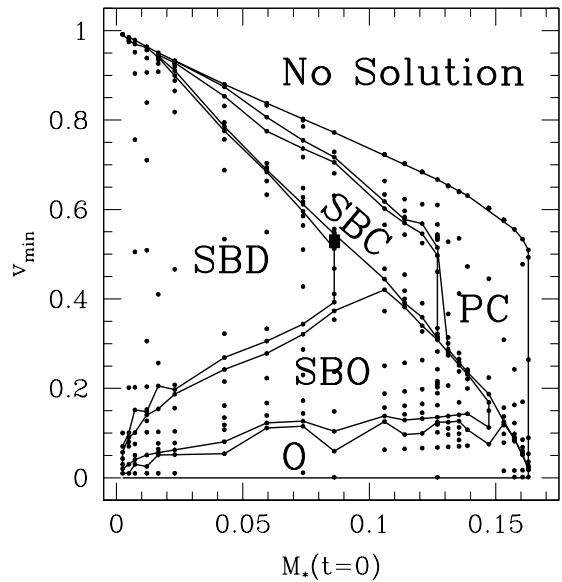}
\caption{Parameter space showing the regions in which various outcomes occur. 
This is the same data shown in Fig.~\ref{fig:pspace} but displayed  with 
respect to the initial magnitude of the velocity perturbation, $v_{\min}$, and 
the initial mass of the star, $M_\star(0)$. Note that $M_\star$ is the gravitational
mass of the \emph{static} star solution and not of the perturbed star.  Since $M_\star(\rho_c)$ is monotonic
in the region we sampled (Fig.~\ref{fig:rho-params}), this figure is essentially a distortion
of Fig.~\ref{fig:pspace}.  The most massive stars shown here have $\rho_c=0.27$ and $M_\star=0.1629$. 
The small black rectangular region located at $(M_\star,v_{\min}) \simeq (0.086,0.53-0.55)$ represents 
a set of solutions that undergo an SBO-type evolution.  
See Fig.~\ref{fig:pspace} caption and text for definition of various phases
that are identified. 
\label{fig:mass-pspace}}
\end{figure}

By sampling $v_{\min}$ and the initial central density 
of the star, $\rho_c$, we have created a type of ``phase diagram'' for
the various ways in which perturbed TOV solutions evolve. 
The phase diagram is shown in Fig.~\ref{fig:pspace}.   
We sample the parameter space by varying the parameter $v_{\min}$ for each value of $\rho_c$. 
Approximately 360 $\{\rho_c,v_{\min}\}$ 
sets were run in order to resolve the phase boundaries. 
Given any combination of the central value of the star's rest-mass density, $\rho_c$, 
and $v_{\min}$, the system will evolve in a fashion specified by the diagram.  In 
Fig.~\ref{fig:mass-pspace}, we display the phases in $(M_\star, v_{\min})$ space.  

The types of dynamical outcomes or ``phases'' identified in 
Figs.~\ref{fig:pspace}--\ref{fig:mass-pspace} are:
\begin{description}
\item[Prompt Collapse (PC):] The initial ``perturbation'' is
so strong that the star is driven directly to black hole formation.  The fluid collapses
homologously---or uniformly---and insignificant amounts of material are ejected before
the black hole forms. 
\item[Shock-Bounce-Collapse (SBC):] The perturbation is not sufficient to spontaneously
form a black hole, but is strong enough to eventually drive the star to collapse. 
The outer part of the star collapses at a faster rate than the interior and eventually bounces 
off the denser core, producing an outgoing shock which expels a significant portion of 
the outer layers of the star. 
\item[Shock-Bounce-Dispersal (SBD):]  This case is quite similar to 
the SBC scenario, except a black hole never forms.  Instead, the 
star contracts until it reaches some maximum density and pressure at the origin which 
is great enough to expel the remainder of the star, leaving behind an ever-decreasing 
amount of matter.  This final explosion results in another outgoing shock wave that 
typically overtakes and engulfs the first shock.  
\item[Shock-Bounce-Oscillation (SBO):] As the perturbation is decreased, the rebound of the 
interior no longer results in complete mass expulsion.  Rather, some matter remains after the 
first two shocks propagate outwards and this matter settles into a new equilibrium state by 
oscillating away any excess kinetic energy via shock-heating.  After the 
oscillations dampen away, a star is left behind that is larger, sparser and hotter than the 
original.
\item[Oscillation (O):] Finally, if the inward velocity is minimal, then the perturbed
star will undergo adiabatic oscillations at its fundamental 
frequency and overtones with a negligible expulsion of mass.
\end{description}
Quantitative definitions and further descriptions of these end states can be found in 
App.~\ref{app:end-states}. 

The phase boundaries---with the possible exception of that 
between the SBO/O states---appear to be quite smooth.  This uniformity lends itself to 
global characterizations, such as a comparison of the dynamical scenarios possible between 
less compact stars (low $\rho_c$) and more compact stars (high $\rho_c$).  
For example, we find that only low $\rho_c$ stars can undergo a complete explosion that 
disperses the star's matter to infinity, and they require
significantly larger perturbations to form black holes.  Both of these aspects 
are intuitive since such stars generate less spacetime curvature.  
On the other hand, more compact stars induce greater spacetime 
curvature, and so are more difficult---and apparently impossible in some cases---to 
completely disperse from the origin.  

From our survey, we have also found that it is not possible to drive some of the less compact stars 
to black hole formation, regardless of the size of the initial velocity perturbation.
Black holes arise through SBC dynamical scenarios for $\rho_c\gtrsim0.007$, and homologous 
collapse to a black hole (PC) only occurs for stars with $\rho_c\gtrsim0.01$.  Since we 
observe Type~II critical phenomena for $0.01\lesssim\rho_c\lesssim 0.05343$ (see 
\cite{noble-choptuik1} for more details), we surmise that arbitrarily small
black holes can form for this entire range of TOV solutions.  For $\rho_c\gtrsim0.05344$, we find 
that the threshold solutions are Type~I solutions, suggesting the smallest black holes that can 
evolve from such stars have finite masses.  
The Type~I behavior seen in perturbed stars will be discussed in Sec.~\ref{sec:type-i-critical}.

In order to compare our results to Novak's, we need to transform our scale to his.  
However, it is unclear what scale Novak used.  He stated masses in terms of solar masses, but 
wrote ``$K=0.1$'' without specifying the units of $K$.  This possibly suggests that he used
geometrized units in that case.  Given this 
uncertainty, we attempt to compare our values to his by determining the $K$ 
that makes the mass of our  last stable TOV solution (i.e. the solution with the maximum mass) 
correspond to his value of $3.16 M_\odot$.  
We will place a ``hat'' over all quantities that we state in these units. 
Using the methods described in App.~\ref{app:unit-conversion}, we find that this
factor of $K$ is 
$\hat{K}=5.42\times 10^5 \mathrm{cm}^{5} \mathrm{g}^{-1} \mathrm{s}^{-2} $.
Let $M_1$ be the mass of the least massive star that can form a black hole through any scenario, and $M_2$ be 
the mass of the least massive star that we observe to undergo prompt collapse to a black hole.   
In our units, we find $M_1 \simeq 0.01656$ at $\rho_c=0.007$, and $M_2 \simeq 0.02309$ at $\rho_c=0.01$. 
Using $\hat{K}$ to convert our masses to his yields $\hat{M}_1=0.320 M_\odot$ versus
$\hat{M}_1=1.155 M_\odot$, and $\hat{M}_2=0.446 M_\odot$ where Novak reports
$\hat{M}_2\approx2.3 M_\odot$.  Note that $\hat{M}_2$ is estimated from Fig.~5 
of \cite{novak}, where a velocity profile equivalent to our $U_2$ 
profile (\ref{v-profile-12}) is used.  Since we have only performed the parameter space survey for $U_1$ 
we cannot say what we would get for $M_2$ when using $U_2$.  However, 
Novak performed a comparison between these two profiles and found that his estimates for 
$M_1$ deviated by about $1\%$ between the two.  Hence, we believe it is adequate 
to quote his result for $U_2$ in order to compare to our result for $U_1$. 

The difference in masses is also obvious in our respective phase diagrams from the parameter space surveys, where 
the point along the $\rho_c$ axis ($n_B$ in Novak's case) at which black holes can
form occurs for noticeably more compact stars in Novak's case~\footnote{Since Novak uses 
$K=0.1$ and since $\rho_c$ scales as $K$, then we may compare our values to his by 
multiplying $0.1$ to his unitless density, $n_B$.}.
Another significant distinction we see in our phase space plot is an absence of 
SBC states for larger $\rho_c$.  Novak seems to observe such scenarios all the way to the 
turnover point ($\rho_c=0.318$), whereas we find that they no longer happen for $\rho_c\gtrsim0.108$. 

Additionally, we believe our study is the first to extensively
cover the subcritical region of NS collapse.  While the method by which
the NSs are perturbed may not be the most physically relevant prescription, we 
are able to see all the collapse scenarios found in  previous  works. 
Much of the previous research focused on compact stars near the turnover point or studied some other 
region exclusively (e.g. \cite{vanriper2}, \cite{vanriper-arnett}, \cite{romero}, 
\cite{font-etal2}, \cite{siebel-font-pap-2001}),
while others individually observed much of the phenomenon without thoroughly scrutinizing the 
boundaries between the phases (\cite{shap-teuk-1980}, \cite{novak}, \cite{gourg2}).  

Our parameter space survey also sheds light on the 
critical behavior observed at the threshold of black hole formation. 
Specifically, we see that the SBD/SBO boundary on the subcritical side of the diagram 
seems to be correlated with the transition from Type~II to Type~I critical behavior.   
The Type~II threshold lies along the SBD/SBC boundary, while the Type~I 
threshold occurs along the line that separates SBO and O cases from black hole-forming cases.  We have 
been able to determine that $\rho_c\approx0.05344$ is the approximate point at which the 
transition from Type~II to Type~I behavior occurs.  For Type~II minimally subcritical solutions near this 
transition,
the matter disperses from the origin but it is difficult to say  if it escapes to infinity since 
our grid refinement procedure is incapable of coarsening the domain.  Consequently, the time step is too 
small to allow for longtime evolutions of these dispersal cases, and we are unable to guarantee that 
they do indeed disperse to infinity.  In addition, the self-similar portion of these marginally subcritical 
solutions entails only a small amount of the original star's matter, the remainder of which 
could, in principle, collapse into a black hole at a time after the inner self-similar component departs from the
origin.  Hence, with our current code, it is difficult to determine the ultimate fate of these 
dispersal scenarios.  

What does this parameter survey suggest about the black hole mass function from driven NS collapse?  
For PC scenarios, the black hole mass is approximately the same as the progenitor star's mass.  The SBC/PC boundary 
marks where the black hole mass function can begin to significantly deviate from the stellar mass function.  The least 
extreme (smallest $v_{\min}$) SBC scenario takes place near $v_{\min}\simeq0.3$, $\rho_c\simeq0.1$ and $M_\star\simeq0.13$. 
Such a large velocity profile may seem unphysical, however, a self-consistent, general relativistic simulation of a 
core collapse supernova in spherical symmetry performed by Liebend\"{o}rfer et al.~\cite{liebendorfer-etal} led to a minimum 
velocity of $\sim-0.6c$ soon after bounce.  This suggests that $v_{\min}\gtrsim 0.3$ is not so unrealistic.  Also, 
it means that Type~II 
behavior may be physically attainable in nature if---in fact---$v_{\min}$ reaches the magnitudes seen in \cite{liebendorfer-etal} 
since $v_{\min}\simeq0.55$ is the smallest velocity profile that leads to Type~II behavior.  
However, we find that $M_\mathrm{BH}$ becomes a power-law only when $v_{\min}$ has been tuned to less than $0.01\%$ 
of the critical value \cite{noble-choptuik1}, suggesting that such cases will not affect the black hole 
mass function significantly.  Unfortunately, we have 
not measured the dependence of $M_\mathrm{BH}$ on $v_{\min}$ and $\rho_c$ in the SBC regime, and---therefore---are not sure
if the distribution is non-trivial.  We hope to measure this in the future.  

Wan et al.~\cite{wan-jin-suen-2008} present a similar phase space survey of a head-on collision between two identical Gaussian distributions of stiff 
matter ($\Gamma=2$) to approximate the head-on merger of identical neutron stars.   The two Gaussian distributions were boosted toward each 
other with the same velocity magnitude.  The amplitude of the boost velocities and the initial central densities of the Gaussian distributions
were varied to explore the nature of the critical surface.   As the central densities were varied, the total baryonic masses of the pulses
were kept constant by adjusting the width of each distribution.  Like us, they find that there is a line that separates black hole forming 
initial conditions
from NS forming conditions.  Unlike our results, however, they find that their line is concave leftward, suggesting that there is a maximum density 
beyond which black hole formation is impossible independent of boost velocity.  
Further, it suggests that at a given initial central density (below 
this upper limit) there are two critical transitions:  from NS-forming to BH-forming to NS-forming---i.e. there is a velocity 
value above which only NS formation is possible.  Wan \cite{wan-thesis-2010} explains further that the second threshold arises because at this point
the merger produces a shock that heats the gas to the point that collapse is prevented.   
In our system, black hole formation is always possible except for the sparsest stars. 

%%%%%%%%%%%%%%%%%%%%%%%%%%%%%%%%%%%%%%%%%%%%%%%%%%%%%%%%%%%%%%%%%%%%%%%%%%%%%%%%%%%%%
%%%%%%%%%%%%%%%%%%%%%%%%%%%%%%%%%%%%%%%%%%%%%%%%%%%%%%%%%%%%%%%%%%%%%%%%%%%%%%%%%%%%%
%%  CHAPTER  %%%%%%%%%%%%%%%%%%%%%%%%%%%%%%%%%%%%%%%%%%%%%%%%%%%%%%%%%%%%%%%%%%%
%%%%%%%%%%%%%%%%%%%%%%%%%%%%%%%%%%%%%%%%%%%%%%%%%%%%%%%%%%%%%%%%%%%%%%%%%%%%%%%%%%%%%
\section{Type~I Critical Phenomena}
\label{sec:type-i-critical}

In this section we describe the Type-I behavior observed when perturbing a 
TOV solution with an imploding pulse of scalar field.  Please see \cite{noble-choptuik1}
for a description of the Type~II phenomena.  

%%%%%%%%%%%%%%%%%%%%%%%%%%%%%%%%%%%%%%%%%%%%%%%%%%%%%%%%%%%%%%%%%%%%%%%%%%%%%%%%%%%%%%%%%%%%%%%%%
%%%%%%%%%%%%%%%%%%%%%%%%%%%%%%%%%%%%%%%%%%%%%%%%%%%%%%%%%%%%%%%%%%%%%%%%%%%%%%%%%%%%%%%%%%%%%%%%%
\subsection{Model Description}
\label{sec:model-description}
As others have done \cite{hawley-choptuik-2000,siebel-font-pap-2001}, we use a minimally-coupled,
massless scalar field to perturb our star solutions dynamically.  The scalar field is 
advantageous 
for several reasons. First, the fact that the two matter models 
are both minimally-coupled to gravity with no explicit interaction between the two ensures that any 
resulting dynamics from the perturbation is entirely due to their gravitational interaction.
Second, the EOM of the scalar field are straightforward to solve numerically and provide little 
overhead to the hydrodynamic simulation.  Third, since gravitational waves cannot exist in spherical symmetry 
and the scalar field only couples to the fluid through gravity, it can serve as a plausible first approximation 
to gravitational radiation acting on these spherical stars.  

We will continue to approximate NSs
as polytropic solutions of the TOV equations with $\Gamma=2$, and the factor in the polytropic
EOS (\ref{polytrope-eos}) will still be set to $K=1$ to keep the system unitless.   Since all stellar 
radii $R_\star$ satisfy $R_\star<1.3$ for such solutions, we will---by default---position the initial scalar field 
pulse at $r=5$.  This has been found to be well outside any star's extent and so ensures that 
the two matter sources are not initially interacting.

%%%%%%%%%%%%%%%%%%%%%%%%%%%%%%%%%%%%%%%%%%%%%%%%%%%%%%%%%%%%%%%%%%%%%%%%%%%%%%%%%%%%%%%%%%%%%%%%%
%%%%%%%%%%%%%%%%%%%%%%%%%%%%%%%%%%%%%%%%%%%%%%%%%%%%%%%%%%%%%%%%%%%%%%%%%%%%%%%%%%%%%%%%%%%%%%%%%
\subsection{The Critical Solutions} 
\label{sec:critical-solutions}

The evolution of the star towards the critical solution and the critical solutions themselves will 
be described in this section.  As the scalar field pulse travels into the star, the star undergoes a
compression phase wherein the exterior implodes at a faster rate than the interior.  This is 
reminiscent of the velocity-induced shock-bounce scenarios described in 
Sec.~\ref{sec:veloc-induc-neutr}.  If the perturbation is weak, then the star will undergo 
oscillations with its fundamental frequency after the scalar field disperses 
through the origin and finally escapes to null infinity (higher harmonics are also excited). 
On the other hand, when the initial scalar 
shell is of  sufficiently large amplitude, the star can be driven to prompt collapse, trapping some of the 
scalar field 
along with the entire star in a black hole.  Somewhere in between, the scalar field can compactify 
the star to a nearly static state that resembles an unstable TOV solution of slightly increased mass.  
The length of time the perturbed star emulates the unstable solution, which we will call the 
\emph{lifetime}, increases as the initial pulse's amplitude is adjusted closer to the critical value, $p^\star$.  
It is expected from this scaling behavior that a perfectly constructed scalar 
field pulse with $p=p^\star$ would perturb the star in such a way that it would resemble the unstable solution 
forever.  This putative, infinitely long-lived state is referred to as the critical solution of the progenitor 
star. 

\begin{figure}[htb]
\includegraphics[scale=0.3]{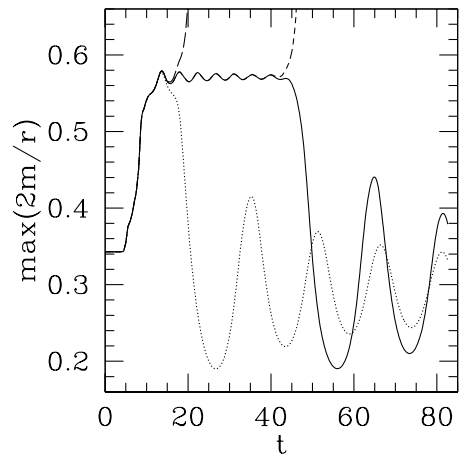}
\caption{Evolutions of ${\max}(2m/r)$ 
from $4$ solutions near 
the critical threshold of a star parameterized by $\Gamma=2$, $\rho_c=0.15$.  Shown are solutions far
from threshold on the supercritical side (long dashes), near threshold on the supercritical side (short 
dashes), near threshold on the subcritical side (solid curve), and far from threshold on the subcritical 
side (dots).  The two solutions near the threshold have  been tuned to within machine precision of the 
critical value.
\label{fig:tuning-maxtmr}}
\end{figure}

\begin{figure}[htb]
\includegraphics[scale=0.3]{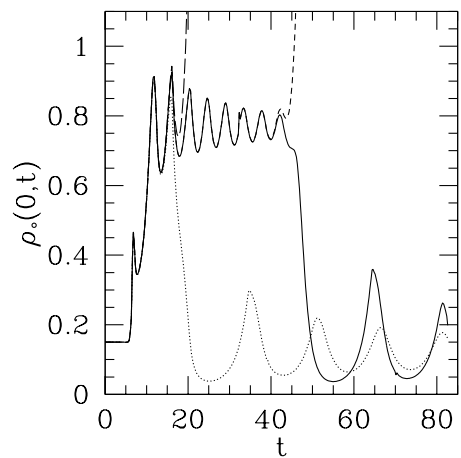}
\caption{
Same as in Fig.~\ref{fig:tuning-maxtmr} but of $\rho_\circ(r\!\!=\!\!0,t)$.
\label{fig:tuning-rhoc}}
\end{figure}

\begin{figure}[htb]
\includegraphics[scale=0.3]{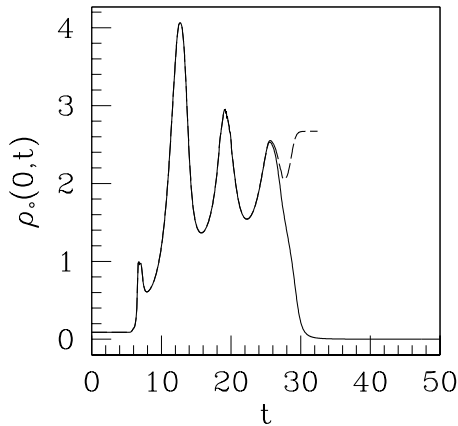}
\caption{Sample evolutions of the 
central rest-mass density for supercritical (dashes) and subcritical 
(solid) solutions from a progenitor star with $\rho_c=0.09$. The solutions have been 
tuned to within machine precision of criticality.  Note that $\rho_\circ(0,t)$
for the supercritical calculation tends to a constant value since the 
``collapse of the lapse'' has effectively frozen the star's evolution near the origin.  
The subcritical solution evolves to an oscillating star that is larger and sparser
than the original state; these oscillations are not visible at the scale used here.
\label{fig:rho-samples-0.09}}
\end{figure}

\begin{figure}[htb]
\includegraphics[scale=0.3]{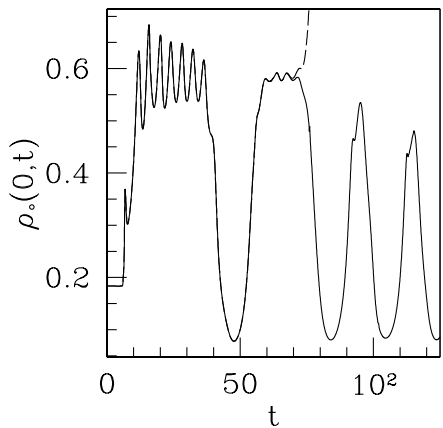}
\caption{
Sample evolutions of the 
central rest-mass density for supercritical (dashes) and subcritical 
(solid) solutions from a progenitor star  with $\rho_c=0.1835$. The $\rho_c=0.1835$ star is the star 
with the smallest initial central density whose nearest-to-critical solution exhibits a momentary departure from the 
unstable equilibrium solution; this is indicated by the break between the two ``plateaus'' in the graph.  
This behavior is seen for most stars above $\rho_c=0.1835$.  
\label{fig:rho-samples-0.1835}}
\end{figure}

Examples of solutions near and far from the critical solution are illustrated in 
Figs.~\ref{fig:tuning-maxtmr}--\ref{fig:tuning-rhoc} for a star with $\rho_c=0.14$.  Here we show the 
evolution of the spatial maximum of $2m/r$, 
${\max}(2m/r)$, and the central density of the star for a series of solutions.  The 
quantity $2m/r$ is, effectively, a measure of the degree of compactification;
the global maximum that $2m/r$ can attain for the static TOV solutions studied herein is approximately $0.61$, and 
$2m/r\rightarrow 1$ when a black hole would form.  The supercritical systems far 
from the threshold quickly collapse to black holes as indicated here by the divergence 
of the central density and compactification factor. 
On the opposite side of the spectrum, we see that subcritical solutions undergo a series of 
oscillations.  The plateau shown 
in the plots represents the period of time during which the marginally subcritical and supercritical 
solutions resemble the critical solution.  We will see shortly that this critical 
solution is actually a star-like configuration oscillating about an unstable TOV solution.  
Interestingly, we do not see the secular growth in the central density with respect to time in these near-threshold 
solutions that others report \cite{radice-2010}; these authors note that it is likely due to their 
use of a multi-dimensional code and lower effective resolution.

\begin{figure}[htb]
\includegraphics[scale=0.3]{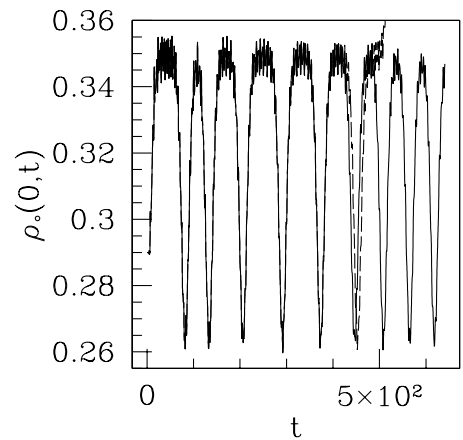}
\caption{
Sample evolutions of the 
central rest-mass density for supercritical (dashes) and subcritical 
(solid) solutions from a progenitor star  with $\rho_c=0.29$.
The supercritical solution undergoes a curious sequence not seen in many 
cases: after it deviates from the subcritical 
solution---instead of collapsing to a black hole immediately---it 
returns to it one last time before collapsing.
\label{fig:rho-samples-0.29}}
\end{figure}

\begin{figure}[htb]
\includegraphics[scale=0.4]{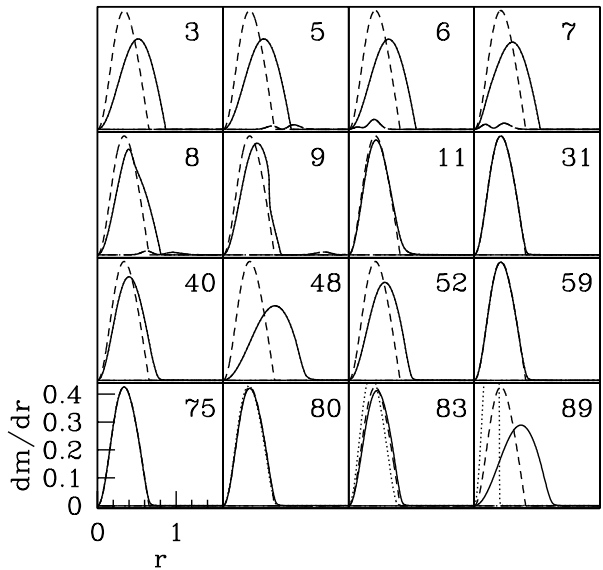}
\caption{Time series of fluid and scalar 
field contributions to $dm/dr$ for the solutions closest to the critical threshold of progenitor stars with $\rho_c=0.197$.
The supercritical (subcritical) fluid contribution is the dotted (solid) curve, and the scalar field 
contribution that gives rise to the supercritical (subcritical) solution is shown 
as a dot-dashed (long-dashed) curve near the bottom of each frame (and is most visible in the third and 
fourth frames).  The $dm/dr$ of the fluid's unstable, equilibrium solution that most closely approximates our
critical solution is shown as the dashed line.  The elapsed proper time measured at spatial infinity of each frame is shown 
in the upper-right corner.  Since the differences between
the supercritical and subcritical scalar field perturbations is on the order of machine precision, the 
subcritical scalar field contribution is completely obscured by the supercritical one.  
Also, the supercritical and subcritical fluid contributions are nearly identical until $t=80$, when
the two solutions begin to diverge from the critical solution. 
\label{fig:dmdr-evol-0.197}}
\end{figure}

Instead of dispersing to spatial infinity as do the solitonic oscillon stars of 
\cite{brady-chambers-goncalves}, the marginally-subcritical 
TOV stars ultimately settle into bound states.  Depending on the magnitude of $p^\star$ for a particular 
progenitor star, the final star solution will either be larger and sparser than the original 
(large $p^\star$), or it will oscillate indefinitely about the original solution. In reality, 
the star will radiate away 
the kinetic energy of the oscillation via some viscous mechanism.  In our model, however, the only 
dissipation is the trivial amount from the numerical scheme, and that from the star shock-heating its
atmosphere---transferring the kinetic energy of the bulk flow into internal energy. 
If the subcritical star settles to a sparser solution, it will 
do this through a series of violent, highly-damped oscillations similar to the SBO scenarios of 
velocity-perturbed stars described in Sec.~\ref{sec:veloc-induc-neutr}.  Examples of such 
subcritical SBO solutions are depicted in 
Figs.~\ref{fig:tuning-maxtmr}--\ref{fig:rho-samples-0.09}.  The damped oscillations are best
illustrated in the marginally subcritical solutions shown in Figs.~\ref{fig:tuning-maxtmr}--\ref{fig:tuning-rhoc}, since 
the oscillations of the subcritical solution of $\rho_c=0.09$ (Fig.~\ref{fig:rho-samples-0.09}) 
occur at an imperceptible scale.  

For these sparser stars, the perturbation required to generate near-critical evolution
is quite large and, consequently, is such that it drives the star 
to significantly \emph{overshoot} the unstable TOV solution, setting it to ring about the unstable 
solution instead.  This metastable ringing decreases with decreasing $p^\star(\rho_c)$, or increasing
$\rho_c$.  For instance, the critical solution of the $\rho_c=0.09$ star seems to correspond to 
an unstable TOV star with central density $\rho_c^\star\simeq2$ that oscillates such that $0<\rho_\circ(0,t)<4$.
The increase in central density---from the initial stable star to the unstable star solution---represents 
an increase by a factor of $22$.
This is to be contrasted with the critical solution for the $\rho_c=0.29$ star which has a central density 
$\rho_c^\star\simeq0.35$---an increase by a factor of $1.2$; this critical solution oscillates 
such that $0.32<\rho_\circ(0,t)<0.38$.  This trend will be discussed further in Sec.~\ref{sec:mass-transf-trans}.

In addition to smaller oscillations about the metastable states for denser initial stars, we see 
from Figs.~\ref{fig:rho-samples-0.1835}--\ref{fig:rho-samples-0.29} that near-critical evolutions
can momentarily depart from their metastable states.  The departures
are illustrated by a break in the plateaus of the $\rho_\circ(0,t)$ distributions.  As $\rho_c$
increases and gets closer to the turnover point, which is located at $\rho_c=0.318$, we see that the 
number of distinct plateaus increases.  The $\rho_c=0.1835$ solution is the smallest initial central 
density where two plateaus are observed, and $\rho_c=0.21$ is the first one where three are seen.  
For higher densities we see an ever-increasing number of plateaus, most likely because the difference 
between the progenitor solution and its corresponding critical solution diminishes.  We explore possible 
causes of these departures in App.~\ref{sec:plateaus}.

As we can see in the time sequence of the scalar field and fluid contributions to $dm/dr$ 
in Fig.~\ref{fig:dmdr-evol-0.197}, the marginally subcritical and supercritical stars leave the 
unstable TOV star configuration only to return to it after one oscillation about the progenitor solution. 
The unstable star solution shown was found by calculating a TOV solution with central density equal to  the time average 
of $\rho_\circ(0,t)$ of the solution tuned nearest to the threshold.  The shock from the outer layers of the star reacting first to 
the increase in curvature is first seen at $t=9$ of this figure.

Making a quantitative comparison of the critical solution to an unstable star is not easy since the 
critical solution is not exactly static.  If we make the assumption that the oscillation is 
sinusoidal, we can take a time-average of the solution when it  most resembles
an unstable star.  We first start with the subcritical solution
that is closest to the threshold.  The periods over which the solution best 
approximates the unstable solution are determined by qualitatively judging where the sequences of 
quasi-normal oscillations begin and end.  The central density, $\rho_c^\star$, of the critical solution 
is then estimated as the time-average of $\rho_\circ(0,t)$ over each of these periods.  
For each system with multiple periods (or plateaus) studied here, we have found the plateau averages all agree
with each other to within their standard deviation.  Hence, we feel that this is a consistent method 
for identifying the unstable star associated with a critical solution.

\begin{figure}[htb]
\includegraphics[scale=0.4]{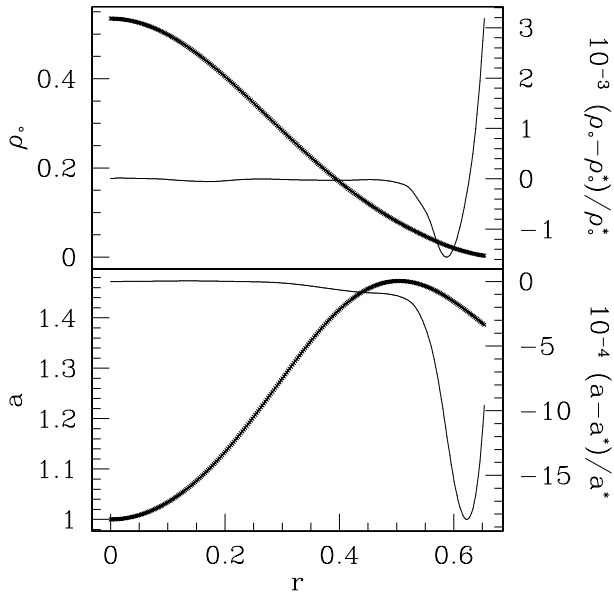}
\caption{
Time-averages ($\times$'s) of $\rho_\circ$ (top) and $a$ (bottom) from the marginally 
subcritical solution compared to those from the 
associated unstable TOV solution ($\rho_\circ^\star$ and $a^\star$, dark solid curves) it best approximates.
The subcritical solution used has been tuned to within machine precision
of the critical solution, and whose initial star has central density of $\rho_c=0.197$.  
Only every eighth point of the time-averaged distributions is displayed. 
Also shown (light solid curves) are the relative differences between these two sets of functions. 
The curves are truncated at the stellar radius of the critical solution. 
\label{fig:type-i-critsol-comparison1}}
\end{figure}

\begin{figure}[htb]
\includegraphics[scale=0.4]{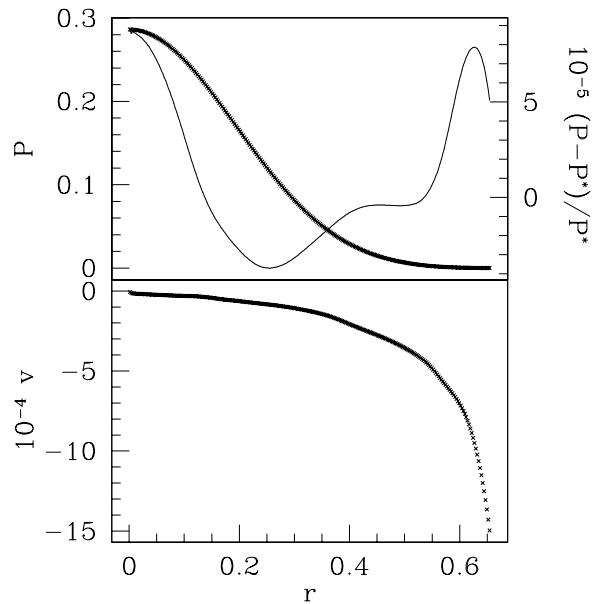}
\caption{
Same as in Fig.~\ref{fig:type-i-critsol-comparison1} but for the functions $P$ (top) and $v$ (bottom).
\label{fig:type-i-critsol-comparison2}}
\end{figure}

After identifying a perturbed star's associated metastable solution, we can compare its shape with 
the solution it oscillates about during a plateau.  To perform this comparison for $\rho_c=0.197$, 
we used  the time-average of the perturbed star during the second plateau and the TOV solution 
with central density $\rho_c^\star$.  The results of this comparison
are shown in 
Figs.~\ref{fig:type-i-critsol-comparison1}--\ref{fig:type-i-critsol-comparison2}, where metric and fluid 
functions from the time-average and the estimated unstable TOV solution are shown together along with their 
differences. These figures clearly show that, during ``plateau epochs'', the critical solution 
closely approximates 
an unstable TOV solution of similar central density.  The relative deviation between the two 
solutions increases near the radius of the star, $R_\star$, which is not surprising since the fluid's time-averaged
velocity is greatest there.  Also, near $R_\star$ the star is most likely interacting with the atmosphere in a 
non-trivial way, which could alter its form near the surface.  In fact, a similar discrepancy 
was observed in the critical boson star solutions in \cite{hawley-choptuik-2000}; they found that 
the critical solutions had a longer ``tail'' than their corresponding static solutions.  
Still, 
the differences we see here are encouraging, and suggest strongly that the critical solutions 
we obtain are perturbed stellar solutions from the unstable branch.

%%%%%%%%%%%%%%%%%%%%%%%%%%%%%%%%%%%%%%%%%%%%%%%%%%%%%%%%%%%%%%%%%%%%%%%%%%%%%%%%%%%%%%%%%%%%%%%%%
%%%%%%%%%%%%%%%%%%%%%%%%%%%%%%%%%%%%%%%%%%%%%%%%%%%%%%%%%%%%%%%%%%%%%%%%%%%%%%%%%%%%%%%%%%%%%%%%%
\subsection{Mass Transfer and the Transition to the Unstable Branch}
\label{sec:mass-transf-trans}

\begin{figure}[htb]
\includegraphics[scale=0.44]{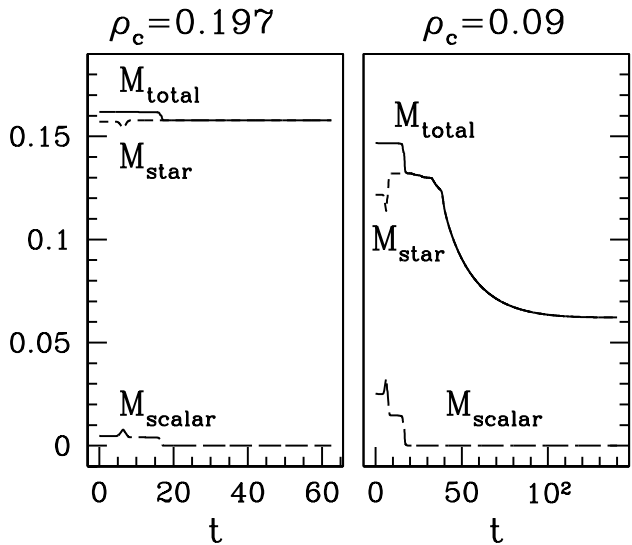}
\caption{Integrated masses of the 
matter fields as a function of time for marginally subcritical
solutions of progenitor stars with $\rho_c=0.197$ (left) and $\rho_c=0.09$ (right). 
The decrease in $M_\mathrm{total}$ (solid curve) at the same time as 
$M_\mathrm{scalar}$ (long dashes) vanishes signifies the scalar field leaving the numerical grid; from the time
it leaves, $M_\mathrm{total}$ is equivalent to $M_\mathrm{fluid}$ (short dashes).
\label{fig:mass-transfer-0.197-0.09}}
\end{figure}

Not only does the perturbing scalar field momentarily increase the spacetime curvature near the origin 
as it implodes through the star, the gravitational interaction of the two matter fields involves
an exchange of mass from the 
scalar field to the star.  
In Fig.~\ref{fig:mass-transfer-0.197-0.09}, we provide a more explicit illustration of the 
mass exchange for two marginally subcritical solutions of stars with $\rho_c=0.197$ and $\rho_c=0.09$.
The total gravitating mass $M_\mathrm{total}$
is calculated via Eq.~(\ref{massaspect}), while $M_\mathrm{fluid}$ ($M_\mathrm{scalar}$) is found 
by integrating $d m_\mathrm{fluid}/dr$
($d m_\mathrm{scalar}/dr$) from the origin to the outer boundary. 
For each case, the non-trivial gravitational interaction of the fluid and scalar field 
can be recognized by the sudden change in their integrated masses, which occurs near $t=7$ in each plot.  
The perturbation for the $\rho_c=0.197$ star is small and does not transfer a significant portion of 
its mass to the star, whereas the perturbation required to drive the $\rho_c=0.09$ 
star to its marginally subcritical state transfers more than a third of its mass to the star.  
This dramatic interaction drives the star to oscillate
wildly about its unstable counterpart---as seen in Fig.~\ref{fig:rho-samples-0.09}---and 
it eventually expels a great deal of the star's mass as it departs from this highly energetic, yet unstable, 
bound state.  The loss of the ejected matter from the grid is clearly 
seen in Fig.~\ref{fig:mass-transfer-0.197-0.09} as 
the long tail of $M_\mathrm{fluid}(t)$, which begins to decrease  well after the scalar field leaves the grid.  

\begin{figure}[htb]
\includegraphics[scale=0.4]{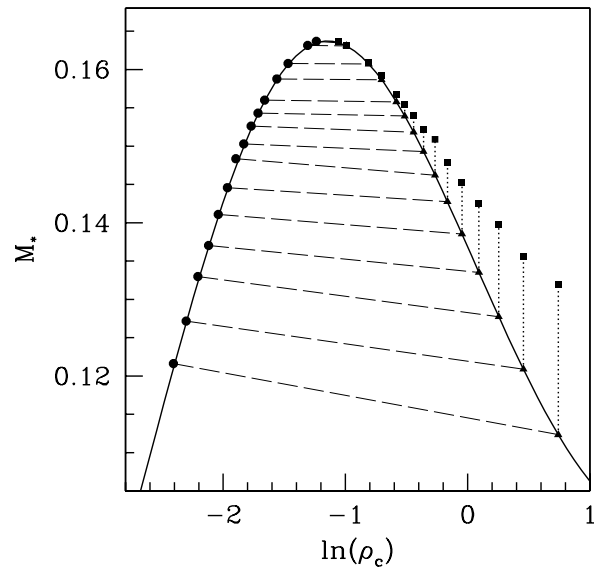}
\caption{Mass versus the log of the central density for 
equilibrium solutions (solid curve),
a few of the initial data sets used (circles), and the critical solutions obtained from 
these initial data sets (triangles and squares).  The central density of a critical solution
was obtained by taking a time average of the central density when the star most resembled the 
attractor solution.  The triangles show where these central densities lie on the unstable
branch, and the mass denoted by a circle or square is of all the fluid in the 
numerical domain.  The dashed and dotted lines indicate the solutions' associations.
\label{fig:mass-crit-sols}}
\end{figure}

To examine how the amount of mass exchange  varies for different critical solutions and to see 
where exactly critical solutions fall on the $M_\star$ versus $\rho_c$ graph of equilibrium solutions, we 
constructed Fig.~\ref{fig:mass-crit-sols}.  The initial star solutions are indicated here on the left
side---the stable branch---while their critical solutions are shown on the right along the unstable
branch.  There are two associated masses for each critical solution: the mass it would have if its profile 
exactly matched the unstable TOV solution with the same time-averaged central density, and its true mass.  Both 
of these masses are indicated in Fig.~\ref{fig:mass-crit-sols} to the right of the turnover point. 
We find that the total fluid mass is always larger than its initial mass, 
whereas the mass of the attractor solution is always \emph{smaller} 
than its stable progenitor.  In addition, as the turnover is approached, both of these deviations diminish 
until, at turnover, the final mass of the fluid distribution corresponds to its initial mass \emph{and} the 
mass of the unstable TOV solution.  

The observed trend that the mass of the unstable TOV solution is always smaller than the progenitor's may 
be explainable in a number of ways.  
First, the assumption that the oscillations of the critical solution about the attractor solution
are sinusoidal would most likely result in overestimates of $\rho_c$ since the oscillations
seem to decay in a nonlinear fashion over time. 
A larger $\rho_c$ would then lead to a  mass estimate smaller than it should be, since 
$dM_\star/d\rho_c<0$ on the unstable branch.  Second, it was seen in 
Figs.~\ref{fig:rho-samples-0.09}--\ref{fig:rho-samples-0.29} that the oscillations 
of the critical solutions decrease with increasing $\rho_c$.  The decrease in the amount of 
energy in these kinetic modes seems to be correlated with the decrease in the exchanged mass.  
A large portion of the exchanged mass must therefore go into the unstable star's kinetic energy.

%%%%%%%%%%%%%%%%%%%%%%%%%%%%%%%%%%%%%%%%%%%%%%%%%%%%%%%%%%%%%%%%%%%%%%%%%%%%%%%%%%%%%%%%%%%%%%%%%
%%%%%%%%%%%%%%%%%%%%%%%%%%%%%%%%%%%%%%%%%%%%%%%%%%%%%%%%%%%%%%%%%%%%%%%%%%%%%%%%%%%%%%%%%%%%%%%%%
\subsection{Type~I Scaling Behavior}
\label{sec:type-i-scaling}

\begin{figure}[htb]
\includegraphics[scale=0.4]{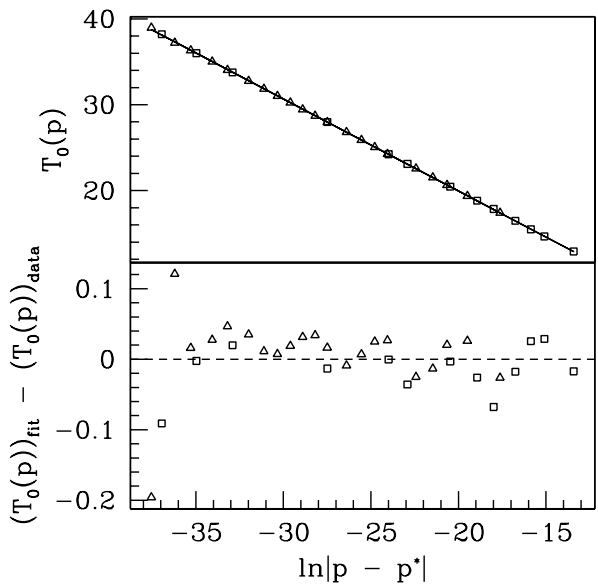}
\caption{(top) Lifetimes, $T_0(p)$, 
for solutions near the threshold that start from a star with $\rho_c=0.14$. (bottom) Deviations of $T_0(p)$
from the best linear fit to the data. 
The scaling exponent, $\sigma$, is found from the negative of the slope of the best linear
fit to  the points.  The fact that both supercritical (triangles) and subcritical (squares) solutions can be used
for calculating $T_0(p)$ is illustrated here by our inclusion of both sets of points. 
The lifetimes shown here are actually those measured at spatial infinity; see the text for further information. 
\label{fig:lifetime-fit}}
\end{figure}

As the amplitude of the initial pulse of scalar field is adjusted 
toward $p^\star$, the lifetime of the metastable, 
near-critical configuration increases.  To quantify the scaling for a given 
initial star solution, the subcritical solution closest to the critical one is first determined.  
This is done by tuning the amplitude of the scalar field pulse, $p$, until consecutive bisections 
yield a change in $p$ smaller than machine precision.  Let $p^\mathrm{lo}$ be the value of $p$
that yields the subcritical solution that most closely approximates the critical solution. 
For each $p$, a unique solution is 
produced that resembles this marginally subcritical solution for different lengths of time, 
determined by how close $p$ is to $p^\star$.  Assuming that the $p^\mathrm{lo}$ solution 
resembles the critical solution longer than any other, the lifetime, $T_0(p)$, is then
the proper time measured at the origin that elapses until $\max\left(2m/r\right)$ deviates
from that of the $p^\mathrm{lo}$ solution by more than $1\%$. These lifetimes  $T_0(p)$ are 
then fit against the expected trend (\ref{type-i-scaling}).  An example of such a fit 
is given in Fig.~\ref{fig:lifetime-fit}.  Since supercritical solutions resemble the critical solution 
as well as subcritical solutions, both kinds can be used when determining the scaling 
exponent $\sigma$.  The exponent is the negative of the slope of the fitted line.  The deviation 
of the code-generated data from the best-fit has an obvious modulation, which may be due to the 
periodic nature of the near-threshold solutions.  Similar modulations in the scaling behavior 
have also been reported for the case of head-on neutron star collisions \cite{kellerman-2010}. 

In practice, the lifetime is determined using the 
proper time elapsed at spatial infinity, $T_\infty$, instead of that measured at the origin.  Let us denote
$\sigma_\infty$ as the scaling exponent measured with $T_\infty$.
In order to get the correct scaling exponent, which would correspond to $1/\omega_{Ly}$ of the 
unstable mode, $\sigma_\infty$ must be rescaled.  Since $T_\infty$ is the same as our coordinate time,
$t$, then 
\beq{
dT_0(t) = \alpha(0,t) \, dt   \quad . \label{propertime-interval}
}
In order to estimate the rescaling factor, we assume that $\alpha(0,t)$ does not vary much when the 
solution is in the near-critical regime, so that 
\beq{
\alpha(0,t) \approx \alpha^\star(0) \quad , \label{propertime-approx}
}
where $\alpha^\star$ is the central value of the lapse of  the unstable TOV solution that corresponds
to the critical solution.  The corrected value of $\sigma$ is then calculated using
\beq{
\sigma = \alpha^\star \sigma_\infty  \quad . \label{sigma-scaling}
}

\begin{figure}[htb]
\includegraphics[scale=0.4]{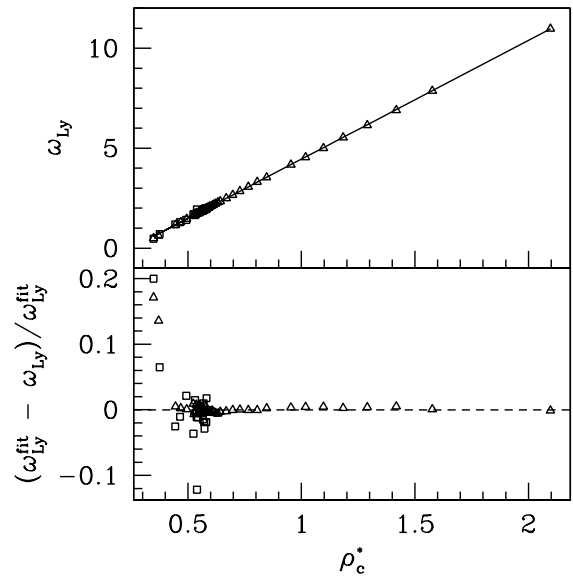}
\caption{(top) Real part of the estimated Lyapunov exponent for a given star solution 
parameterized by $\rho_c^\star$ using the first plateau (triangles) and the second plateau (squares).  
${\max}(2m/r)$ was used to calculate the $\omega_{Ly}$
shown here. (bottom) The relative deviation  of the data from the best linear
fit to data from the first plateau (\ref{sigma-rhoc-fit}). 
\label{fig:type-i-scaling}}
\end{figure}

We have performed fits for $\sigma_\infty$ and then rescaled them using the above procedure to obtain 
an estimate of $\sigma$
for $55$ different initial TOV stars.  The Lyapunov exponent for a critical solution is 
$\omega_{Ly} = 1/\sigma$.   The variation of $\omega_{Ly}$ with $\rho_c^\star$ 
is shown in Fig.~\ref{fig:type-i-scaling}.  We find that $\omega_{Ly}(\rho_c^\star)$
is fit surprisingly well by the linear relationship
\beq{
\omega_{Ly} = 5.93 \rho_c^\star - 1.475  \quad . \label{sigma-rhoc-fit}
}

In order to verify that the calculated $\sigma$ values are, indeed, equal to $1/\omega_{Ly}$, we
need to calculate the fundamental modes of the unstable star solutions.  To the extent of 
the authors' knowledge and that of others \cite{lindblom,sterg}, this has not been done for the particular 
EOS used.  However, the equations governing radial pulsations of stars in general relativity are well-known
(see \cite{mtw,btm} and references therein).  Our method for their solution follows ``Method 1-A'' 
of \cite{btm}, which exploits the fact that the equation to be solved has the Sturm-Liouville form.
Since the fundamental mode, $\omega_0$, of these unstable star solutions is expected to be the unstable mode that
we tune away, we expect $\omega_0=\omega_{Ly}$.  For each unstable star with 
$\rho_\circ=\rho_c^\star$, we calculate $\omega_0$ by iteratively integrating the eigenfunctions in 
first-order form from $r=0$ to $r=R_\star$.  After each iteration, we lower (raise) our guess for $\omega_0$
depending on whether the solution has one (zero) nodes. This bisection process proceeds until we have 
found $\omega_0$ to at least six-digits.  

\begin{figure}[htb]
\includegraphics[scale=0.4]{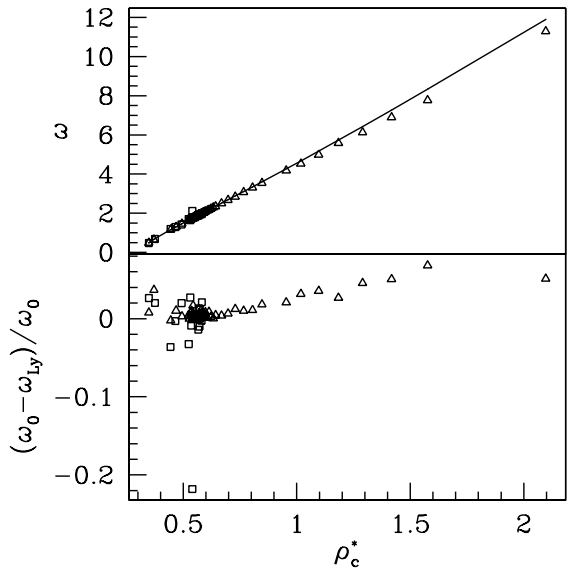}
\caption{(top) Comparison of $\omega_{Ly}$ from the first plateau data (triangles)
and $\omega_{Ly}$ from the second plateau data  (squares) to 
the fundamental mode frequencies, $\omega_0$, of the corresponding unstable TOV solutions (connected dots).
(bottom) Relative deviations between $\omega_{Ly}$ and $\omega_0$. 
\label{fig:normal-modes}}
\end{figure}

A comparison between $\omega_0$ and $\omega_{Ly}$ is shown in Fig.~\ref{fig:normal-modes}. 
The Lyapunov exponents deviate from the fundamental mode frequencies of unstable solutions by no 
more than $7\%$ for all $\rho_c^\star$. 
The relative difference, $\left(\omega_0-\omega_{Ly}\right)/\omega_0$, 
is seen to grow with $\rho_c^\star$.  This may be explained by the possible bias we mentioned 
earlier in how we calculate $\rho_c^\star$ from the near-threshold solution.
Stars with larger $\rho_c^\star$ oscillate with a larger amplitude that tends to decay with 
time (e.g. Fig.~\ref{fig:rho-samples-0.09}).  If one were to assume this decay is the result of 
the threshold solution shedding excess kinetic energy, then our time-averages of 
$\rho_\circ(r=0)$ would yield excessively large $\rho_c^\star$ values. 

Before leaving this section, we wish to comment on the  universality of our system's critical solutions.  
Unlike systems with one unstable static solution---as seen in the Einstein-Yang-Mills model \cite{choptuik-chmaj-bizon,choptuik-hirshmann-marsa}, for example---the TOV 
system admits a family of static critical solutions (i.e. the unstable branch of TOV solutions).  
As demonstrated in \cite{liebling-lehner-neilsen-palenzuela-2010,jin-suen-2006,wan-jin-suen-2008,wan-submitted-2010,wan-thesis-2010}, one can 
perturb \emph{unstable} TOV  solutions in a number of ways to demonstrate Type~I behavior; in this kind of method, one starts with an 
unstable solution and tunes it to the critical solution by adjusting a parameter of the initial data that 
acts to eliminate the single unstable mode.  
In our study we demonstrate that Type~I behavior 
of TOV solutions can also be found when one starts from a \emph{stable} solution and tunes the perturbing agent that drives the star 
to the unstable branch.   We have demonstrated that, at least for the perturbing methods we have explored, the mapping from stable to unstable 
solutions followed no obvious trend.   We therefore cannot predict what critical solution a particular set of initial data will 
tend toward. On the other hand, the calculations performed by the others begin with initial data 
very near an unstable solution that is ultimately identified with a critical solution. 
These computations demonstrate that the unstable branch serves as a family of 1-mode unstable solutions, whereas our method additionally demonstrates
that the unstable branch is the family of 1-mode unstable solutions to which stable solutions are 
attracted---at least for the scenarios we examined. 

%%%%%%%%%%%%%%%%%%%%%%%%%%%%%%%%%%%%%%%%%%%%%%%%%%%%%%%%%%%%%%%%%%%%%%%%%%%%%%%%%%%%%
%%%%%%%%%%%%%%%%%%%%%%%%%%%%%%%%%%%%%%%%%%%%%%%%%%%%%%%%%%%%%%%%%%%%%%%%%%%%%%%%%%%%%
%%  CHAPTER  %%%%%%%%%%%%%%%%%%%%%%%%%%%%%%%%%%%%%%%%%%%%%%%%%%%%%%%%%%%%%%%%%%%
%%%%%%%%%%%%%%%%%%%%%%%%%%%%%%%%%%%%%%%%%%%%%%%%%%%%%%%%%%%%%%%%%%%%%%%%%%%%%%%%%%%%%
\section{Conclusion}
\label{sec:concl-future-work}

In this work, we simulated spherically-symmetric relativistic perfect fluid flow in the strong-field regime of 
general relativity.  Specifically, a perfect fluid that admits a length scale, for example one that 
follows a relativistic ideal gas law,  was used to investigate the dynamics of 
compact, stellar objects.  A stiff equation 
of state was used to approximate the behavior of some realistic state equations for NS matter.  
The stars served as initial data for a parameter survey, in which we drove them to 
collapse using either an initial velocity profile or a pulse of massless scalar field.   Both types of 
critical phenomena were observed using each of the two mechanisms.  The parameter space survey 
provided a description of the boundary between Type~I  and Type~II behavior, and illustrated the 
wide range of dynamical scenarios involved in stellar collapse.  We found that the non-black hole 
end states of solutions near the threshold of black hole seemed to be correlated to the 
type of critical behavior observed.  For instance, Type~I behavior seemed to always entail 
subcritical end states that were bound and star-like.  Type~II behavior, 
on the other hand, was observed to coincide with dispersal end states.

Since the unstable branch of TOV solutions has been known for decades, many 
anticipated that TOV solutions would exhibit some kind of Type~I behavior. 
This paper describes the first in depth analysis
of Type~I phenomena associated with hydrostatic solutions in that the Lyapunov exponents of the critical solutions were measured for 
a variety of cases.  
We  verified that the Lyapunov exponents agree well with the normal mode 
frequency of their associated unstable TOV solutions, confirming that the critical solutions are TOV solutions
on the unstable branch.
The exponents were found to follow a linear relationship as a function of the time-averaged central densities of their associated
critical solutions.  We also discovered that the Type~I critical solutions 
coincided with perturbed unstable hydrostatic solutions which were typically more massive than their
progenitor stars.  

In the future, we hope to address a great number of topics that expand on this work.  
First, the supercritical section of parameter space demands further exploration in order to investigate 
how much matter can realistically be ejected from shock/bounce/collapse scenarios.  In addition, the 
ability to follow spacetimes after the formation of an apparent horizon would allow us to study the 
possible simultaneous overlap of Type~I and Type~II behavior.  
It would also allow us to measure the ultimate mass distribution of black holes, as we are 
able only to measure the black hole masses at the time of formation which neglects any 
subsequent mass accretion.  
Ultimately, it is our goal to 
expand the model a great deal, making the matter description more realistic and eliminating 
symmetry.  As a first step, we wish to develop adaptive mesh refinement procedures 
for conservative systems that will be  required to study critical phenomena of stellar 
objects in axial-symmetry \cite{choptuik-etal-2003}.  Also, we wish to 
examine how Type~II behavior changes in the context of realistic equations of state.  For example, 
realistic equations of state effectively make the adiabatic index of the fluid a function of the fluid's
density and temperature, and, to date, critical behavior in perfect fluids has only been described for 
fluids with constant adiabatic index.

%%%%%%%%%%%%%%%%%%%%%%%%%%%%%%%%%%%%%%%%%%%%%%%%%%%%%%
%%%%%%%%%%%%%%%%%%%%%%%%%%%%%%%%%%%%%%%%%%%%%%%%%%%%%%

\begin{acknowledgments}

The authors wish to acknowledge financial support from CIFAR, NSERC, NSF PHY 02-05155, and NSF CDI AST 10-28087.  
SCN wishes to thank I. Olabarrieta for many helpful comments.  
All numerical calculations were performed on UBC's vn PIII and vnp4 clusters (supported by 
CFI and BCKDF).

\end{acknowledgments}

%%%%%%%%%%%%%%%%%%%%%%%%%%%%%%%%%%%%%%%%%%%%%%%%%%%%%%%%%%%%%%%%%%%%
%%%%%%%%%%%%%%%%%%%%%%%%%%%%%%%%%%%%%%%%%%%%%%%%%%%%%%%%%%%%%%%%%%%%
%%%%%%%%%%%%%%%%%%%%%%%%%%%%%%%%%%%%%%%%%%%%%%%%%%%%%%%%%%%%%%%%%%%%
%   APPENDIX   %%%%%%%%%%%%%%%%%%%%%%%%%%%%%%%%%%%%%%%%%%%%%%%%%%%%%%
%%%%%%%%%%%%%%%%%%%%%%%%%%%%%%%%%%%%%%%%%%%%%%%%%%%%%%%%%%%%%%%%%%%%
%%%%%%%%%%%%%%%%%%%%%%%%%%%%%%%%%%%%%%%%%%%%%%%%%%%%%%%%%%%%%%%%%%%%
\appendix

%%%%%%%%%%%%%%%%%%%%%%%%%%%%%%%%%%%%%%%%%%%%%%%%%%%%%%%%%%%%%%%%%%%%
%%  APPENDIX A  %%%%%%%%%%%%%%%%%%%%%%%%%%%%%%%%%%%%%%%%%%%%%%%%%%%%%%%%%%%
%%%%%%%%%%%%%%%%%%%%%%%%%%%%%%%%%%%%%%%%%%%%%%%%%%%%%%%%%%%%%%%%%%%%
\section{Conversion of Units and Scale}
\label{app:unit-conversion}

When theoretical calculations are made in the theory of general relativity, it is customary 
to use ``geometrized units'' in which $G = c = 1$ (see App.~E of 
\cite{wald} for a comprehensive discussion on the conversion to and from 
geometrized units, only a few key ideas will be mentioned here). 
In such units, scales or dimensions of 
mass ($M$) and time ($T$) are transformed into scales of length ($L$) only, by multiplying 
by appropriate factors of $G$ and $c$.  For instance, because of how 
$G$ and $c$ scale with mass and time, one can easily derive that a quantity $\mathcal{Q}$
that scales like $L^lM^mT^t$, can be converted into geometrized units by 
multiplication of $c^t \left(G/c^2\right)^m$.  After the conversion to 
geometrized units, $\mathcal{Q}$  scales as $L^{l+m+t}$.  

Since the equations governing the ultra-relativistic fluid are all invariant
under changes in the fundamental length scale $L$, such fluids naturally follow 
self-similar behavior \cite{cahill-taub}.  The inclusion of $\rho_\circ$ in the 
system eliminates this intrinsic scale-invariance via the EOS.  For example, 
when using the polytropic EOS, $P = K \rho_\circ^\Gamma$, 
the constant $K$ has dimensions $L^{2\left(\Gamma - 1\right)}$ in geometrized units and 
$L^{3\Gamma-1} M^{1-\Gamma} T^{-2}$ in non-geometrized units. 
Hence, one may set the fundamental length scale of the system
by choosing a value for $K$ \cite{cook-shap-teuk-1992,cook-shap-teuk-1994}.
Since all physical quantities are expressible in dimensions of $L$ in geometrized 
units, the quantities of static \emph{and} dynamic systems 
which use one set $\{K,\Gamma\}$ should be exactly the same as 
those using another set $\{\hat{K},\hat{\Gamma}\}$,
modulo a rescaling of each quantity by the factor  
\beq{
\left(\hat{L}/L\right)^n = 
\left({\hat{K}}^{1/2\left(\hat{\Gamma}-1\right)} / {K}^{1/2\left(\Gamma-1\right)}\right)^n
\quad , \label{rescaling-factor}
}
where $n$ depends on how the particular quantity scales with length.
Thus, setting $K=1$ makes the 
system dimensionless, and this is the approach used in the paper.  This choice simplifies
the comparison of two solutions having different values of $K$ and $\Gamma$.

In order to transform from our dimensionless system to one with dimensions, one must 
first set the scale by fixing $K$.  Let $\hat{X}$ be a quantity that has 
dimensions of $L^lM^mT^t$, and $X$ be the corresponding dimensionless quantity.  In order to 
transform $X$ into $\hat{X}$, one may use the following equation
\beq{
\hat{X} = K^x c^y G^z X \quad , 
\label{unit-transform}
}
where 
\beq{
x = \frac{l + m + t }{ 2 \left( \Gamma - 1 \right)}  
\quad , \quad 
z = - \frac{ l + 3 m + t }{2}
\label{unit-conv-exponents1}
}
\beq{
y = \frac{ \left( \Gamma - 2 \right) l + \left( 3 \Gamma - 4 \right) m  - t }{\Gamma - 1}
\quad . \label{unit-conv-exponents2}
}

When presenting results of TOV solutions using polytropic state equations, it is 
customary to choose $K$ in such a way that the maximum stable mass for 
the given polytrope corresponds to that of the Chandrasekhar mass, $1.4 M_\odot$. 
As an example, a mass $\hat{M}(K)$ expressed in units can be 
calculated from the 
dimensionless $M(K=1)$ via the above formula (since $\hat{M}$ has dimensions of only mass, then
$l=0, m=1, t=0$):
\beq{
\hat{M}(K)  =  K^{1/2\left(\Gamma-1\right)} c^3 c^{-1/\left(\Gamma-1\right)} G^{-3/2} 
               M(K=1) \quad .    \label{mass-units}
}

Since the TOV solutions for $\Gamma=2$ and $K=1$ yield a maximum stable mass of 
$0.164$, then the $K$ that would make $\hat{M}(K) = 1.4 M_\odot$ would be approximately 
$10^5 \mathrm{cm}^{5} \mathrm{g}^{-1} \mathrm{s}^{-2}$, in cgs units.  
The radius of this maximum mass star is $0.768$ with $K=1$, and is 
about $9.4\,\mathrm{km}$ with $K=10^5 \mathrm{cm}^{5} \mathrm{g}^{-1} \mathrm{s}^{-2}$.  

%%%%%%%%%%%%%%%%%%%%%%%%%%%%%%%%%%%%%%%%%%%%%%%%%%%%%%%%%%%%%%%%%%%%%%%%%%%%%%%%%%%%%
%%%%%%%%%%%%%%%%%%%%%%%%%%%%%%%%%%%%%%%%%%%%%%%%%%%%%%%%%%%%%%%%%%%%%%%%%%%%%%%%%%%%%
%%%%%%%%%%%%%%%%%%%%%%%%%%%%%%%%%%%%%%%%%%%%%%%%%%%%%%%%%%%%%%%%%%%%
%%  APPENDIX B  %%%%%%%%%%%%%%%%%%%%%%%%%%%%%%%%%%%%%%%%%%%%%%%%%%%%%%%%%%
%%%%%%%%%%%%%%%%%%%%%%%%%%%%%%%%%%%%%%%%%%%%%%%%%%%%%%%%%%%%%%%%%%%%
\section{Calculating the Initial Star Solution with an Ingoing Coordinate Velocity}
\label{sec:calc-init-star}

Initializing the star with a certain coordinate 
velocity instead of the \emph{Eulerian} velocity, 
$v= a U / \alpha$, 
couples the Hamiltonian constraint (\ref{polar-areal-hamiltonian-const}) and the slicing condition 
(\ref{polar-areal-slicing-condition}) by introducing $\alpha$ and $a$ into their right-hand sides.  
In order to explicitly show how the right-hand sides  change, the conserved variables
must be expressed in terms of the coordinate velocity and primitive variables 
via Eqs.~(\ref{D}-\ref{Phi}):
\beq{  
\frac{a '}{a} = a^2 \left\{ 4 \pi r \left[ 
\frac{\rho_\circ h}{1 - \left(\frac{a U}{\alpha}\right)^2} - P \right] - \frac{1}{2 r^2} \right\} 
+ \frac{1}{2 r^2}
\quad , \label{vp-hamconstraint}
}
\beq{
\frac{\alpha '}{\alpha}  
= a^2 \left\{ 4 \pi r \left[ \rho_\circ h 
\frac{ \left(a U / \alpha\right)^2}{1 - \left(\frac{a U}{\alpha}\right)^2}
  + P \right] + \frac{1}{2 r^2} \right\}  - \frac{1}{2 r^2} 
\quad . \label{vp-slicing-condition} 
}

The coupling of these equations complicates their numerical solution.  We will 
briefly describe how they are solved here.  We start by solving the TOV 
equations, adjusting the lapse so that $\alpha a |_{r=r_{\max}} = 1$. 
Given $U_\circ$, $U(r)$ is specified via Eq.~(\ref{v-profile-12}), and
$\alpha,a$ are re-calculated via a 2-dimensional Newton-Raphson 
method which solves Eqs.~(\ref{vp-hamconstraint}-\ref{vp-slicing-condition}) at each grid point.  The 
integration starts at the origin with $\alpha(r=0),a(r=0)$  from the TOV solution. 
The Eulerian velocity, $v = U a / \alpha$, is calculated using $\alpha,a$ at this stage. 
Since the parameterization for $\alpha$ is chosen at the origin, the 
outer boundary condition, $\alpha a |_{r=r_{\max}} = 1$, will not necessarily be satisfied.  
In order to impose this outer boundary condition and calculate the 
final values of $\alpha(r)$ and $a(r)$, the \emph{uncoupled}
Hamiltonian constraint (\ref{polar-areal-hamiltonian-const}) and 
slicing condition (\ref{polar-areal-slicing-condition}) 
are then solved using the $v$ calculated in the previous step.  

%%%%%%%%%%%%%%%%%%%%%%%%%%%%%%%%%%%%%%%%%%%%%%%%%%%%%%%%%%%%%%%%%%%%%%%%%%%%%%%%%%%%%
%%%%%%%%%%%%%%%%%%%%%%%%%%%%%%%%%%%%%%%%%%%%%%%%%%%%%%%%%%%%%%%%%%%%%%%%%%%%%%%%%%%%%
%%%%%%%%%%%%%%%%%%%%%%%%%%%%%%%%%%%%%%%%%%%%%%%%%%%%%%%%%%%%%%%%%%%%
%%  APPENDIX C  %%%%%%%%%%%%%%%%%%%%%%%%%%%%%%%%%%%%%%%%%%%%%%%%%%%%%%%%%%
%%%%%%%%%%%%%%%%%%%%%%%%%%%%%%%%%%%%%%%%%%%%%%%%%%%%%%%%%%%%%%%%%%%%
\section{Perturbed Neutron Star End States}
\label{app:end-states}

Differentiating between some of the types of outcomes is difficult.  
To aid in this process, we examined how various quantities varied with time at the star's 
radius, $R_\star(t)$.  We define $R_\star(0)$ as the radius of the last numerical 
cell before which $\rho_\circ$ falls below the 
floor density~\footnote{As do many computational fluid dynamics codes, our numerical 
scheme prevents non-positive density and pressure values from arising by imposing
minimum, positive values on these quantities. We call this the ``floor'' or ``floor state.'' 
See \cite{noble-choptuik1} for further details.}, and set
$R_\star(t)$ to be the radius at which $\rho_\circ(r,t) = \rho_\circ(R_\star(0),0)$ to within
some finite precision.  This served as a fair approximation to the worldline of the fluid element 
originally  at $R_\star(0)$, however, we do not assume that $R_\star(t)$ is that of a Lagrangian observer.
The Eulerian velocity at $r=R_\star(t)$ is also considered and will be referred to as $v_\star$. 

The boundary between SBO and O outcomes may be the most imprecisely determined one. 
This is due to the fact that the shock in SBO cases weakens as the 
perturbation is reduced, making it difficult to tell if a bounce actually happens and whether 
the subsequent oscillations take place about a different star solution.  In addition,
an O system may form a minor shock at first, but still maintain nearly-constant amplitude 
oscillations, indicating the absence of significant shock-heating.  Herein, an O state
is defined as a star which lost less than $1\%$ of its mass over the first six periods of its
fundamental mode of oscillation.  This choice of cutoff is motivated by two facts:  1) evolutions which 
seem to be oscillating about the initial solution still lose mass, because the oscillations 
still eject minute amounts of matter from the star's surface;  2) those evolutions which 
are obviously SBO seem to eject most of the expelled matter within the first 6 oscillations. 
Using this definition, we estimate the systematic error of the SBO/O boundary to be no larger than 
$0.05$ in $v_{\min}$.

%\clearpage 
\begin{figure}[htb]
\includegraphics[scale=0.42]{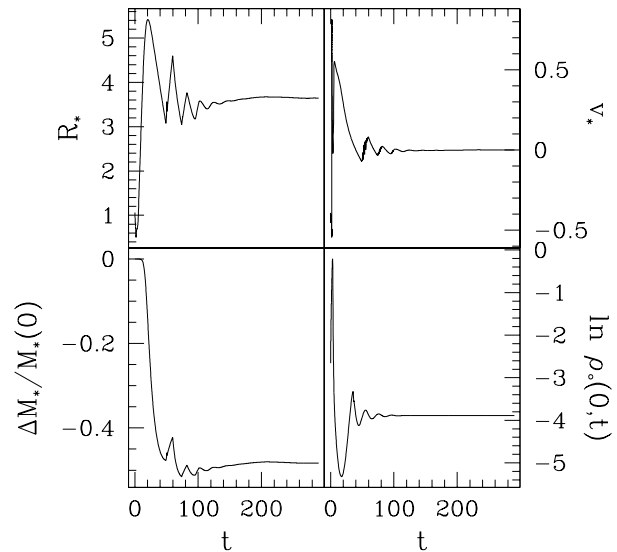}
\caption{Evolutions of stellar 
radius ($R_\star$), velocity at $R_\star$ ($v_\star$), relative stellar mass 
deviation from initial time ($\Delta M_\star(t)/M_\star(0)$), and the natural logarithm of the central density 
for a SBO case.  The defining parameters for this run are 
$\rho_\circ(0,0) = 0.02$, $v_{\min}(0) = 0.397$, $M_\star(0) = 0.1185$. 
\label{fig:sb-oscil}}
\end{figure}

\begin{figure}[htb]
\includegraphics[scale=0.4]{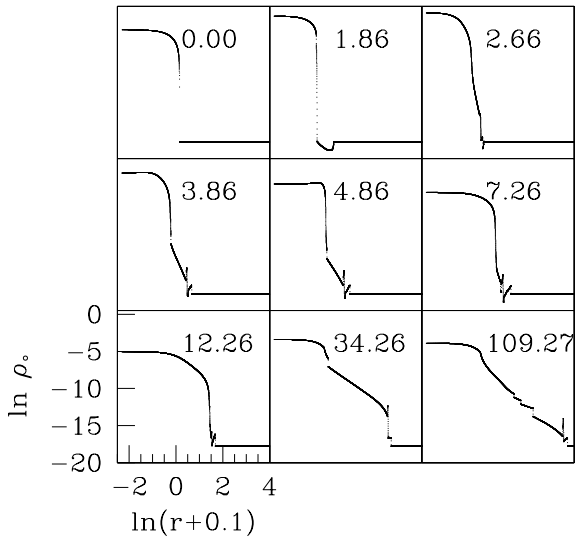}
\caption{Time sequence of $\ln{\rho_\circ(r,t)}$ 
versus $\ln(r+0.1)$ for the same SBO scenario shown 
in Fig.~\ref{fig:sb-oscil}.  
\label{fig:lnrho-sb-oscil}}
\end{figure}

Histories of the star's radius, change in mass, central density and velocity at its outer 
edge for a case that epitomizes an SBO state are plotted in Fig.~\ref{fig:sb-oscil}.  
The star first undergoes a quick shock and bounce at its edge which seems to play an insignificant role 
in the subsequent evolution.  This is indicated by the first maxima in $v_\star$  near $t \approx 3.2$. 
While the shock propagates out of the star, the inner part of the star continues to infall and 
rebounds from the origin, which is responsible for ejecting the majority of the matter from the star.  
The apex of the rebound takes place near $t=10$,
when the star reaches minimum size and maximum central density, and when the star begins to lose 
a significant portion of its initial mass---up to $43\%$ in total.  This large change in $M_\star$ signifies
how poorly $R_\star(t)$ follows the path of a Lagrangian observer in this case; however, 
we still feel tracking quantities along this path produces information with which we can consistently 
differentiate end states.  In order to illustrate how the SBO star's distribution of mass 
changes with time, we show snapshots of $\rho_\circ(r,t)$  in Fig.~\ref{fig:lnrho-sb-oscil}.
The initial shock ($t\simeq1.86$) and bounce ($t\simeq2.66$) are clearly 
seen early on in the time sequence, while the subsequent rebounds 
of the interior are seen later in time.  One can also see that the first 
rebound of the core ($2.66 \lesssim t \lesssim 3.86$) is responsible for most of the ejection of matter, even though the 
initial bounce near the star's surface involves the strongest shock.  
The ensuing oscillations after $t\simeq10$ are evident in all the quantities shown.  
The star finally settles to a time-independent state with a smaller central density, 
larger radius and smaller mass than it had initially.  

\begin{figure}[htb]
\includegraphics[scale=0.42]{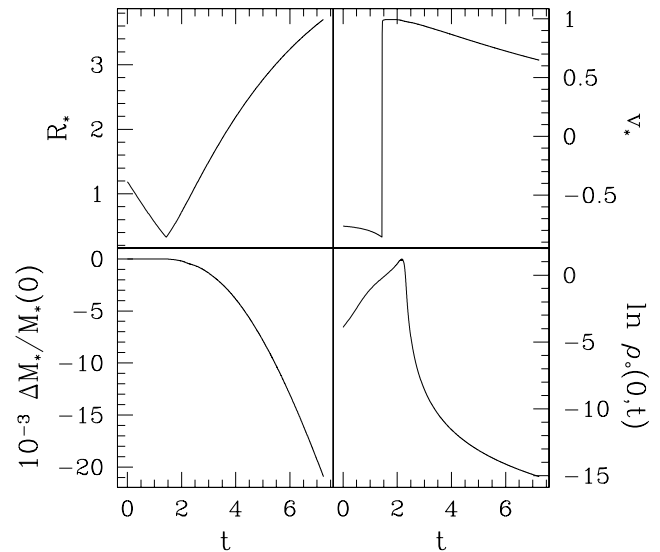}
\caption{Evolutions of stellar radius ($R_\star$), velocity at $R_\star$ ($v_\star$), relative stellar mass 
deviation from initial time ($\Delta M_\star(t)/M_\star(0)$), and the natural logarithm of the central density 
for a SBD star.  The defining parameters for this run are 
$\rho_\circ(0,0) = 0.02$, $M_\star(0) = 0.0726$, $R_\star(0)=1.1885$, and $v_{\min}(0) = 0.766$.
\label{fig:sb-disp}}
\end{figure}

\begin{figure}[htb]
\includegraphics[scale=0.4]{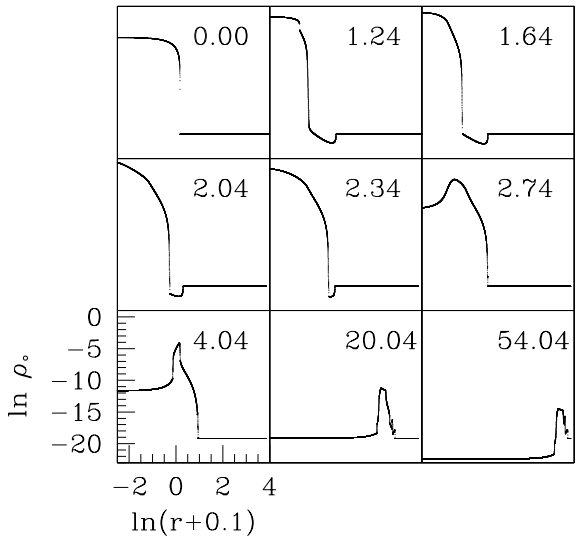}
\caption{Time sequence of 
$\ln{\rho_\circ(r,t)}$ versus $\ln(r+0.1)$ for the same 
SBD scenario shown in Fig.~\ref{fig:sb-disp}.  By $t=54.04$, $\rho_\circ$ has fallen well 
below the floor's density in the vicinity of the origin.
\label{fig:lnrho-sb-disp}}
\end{figure}

It is also sometimes difficult differentiating SBO states from SBD states since 
perturbed stars with smaller $\rho_c$ on the SBD side near the SBD/SBO boundary often homologously 
inflate to arbitrary sizes.  The central densities of these stars diminish to magnitudes comparable to the 
floor density.
In contrast the denser stars close 
to the SBC/SBD border tend to disperse completely from the origin in a shell of matter that has
compact support.  In order to ensure that these ``inflated'' stars will not ultimately 
settle into a new equilibrium configuration, we typically let the evolution last until 
the central density of the distribution becomes comparable to the floor density
and increase the size of the grid to accommodate for the expansion.  If, at this time, 
$v(r) > 0$ for all $r$ and $d \rho_\circ(0,t) / dt < 0 $ are still satisfied, then the particular 
case is labeled as a dispersal, or SBD variety. 
An archetypal example of an SBD case involving a compact star is shown in 
Figs.~\ref{fig:sb-disp}--\ref{fig:lnrho-sb-disp}.

The small rectangle near the upper-right corner of the SBD region in 
Figs.~\ref{fig:pspace}--\ref{fig:mass-pspace} represent 3 runs with $\rho_c=0.05$
that exhibited SBO behavior.  It remains
to be seen whether or not these cases are dominated by numerical artifacts---that is, the 
remnant star may converge away as $\Delta r \rightarrow 0$---or, if they instead 
represent the sparsest instances of SBD type evolutions along the black hole threshold line. 
If they are real solutions, then each section of the parameter space diagram may not be as 
homogeneous as illustrated here.  Interestingly, these 3 runs are near the region where the 
black hole threshold behavior changes from being of Type~II to Type~I ($\rho_c\approx0.05344$).

Since our choice of coordinates (\ref{metric}) precludes a black hole from forming in finite time, 
we need a fairly rigorous prescription for \emph{predicting} when they would form.
Empirically, we have found that those systems which attain 
${\max}(2m/r) > 0.7$ will asymptote to a state that resembles a black hole in 
our coordinates---where $a$ diverges and $\alpha$ shrinks to an exponentially small magnitude
at the origin.  These all provide strong evidence that the simulated spacetime contains a black hole.
If all goes well, we label any spacetime that reaches ${\max}(2m/r) > 0.995$ 
a ``black hole''.
Since such spacetimes involve extremely steep gradients, it is often difficult to stably integrate
the equations of motion until this threshold is achieved.  Consequently we assume that any evolution, 
which crashes and satisfies ${\max}(2m/r) > 0.7$, will eventually give rise to a black hole.  
Otherwise, the system is assumed to be one without a black hole and is either of type O, SBO or SBD. 

\begin{figure}[htb]
\includegraphics[scale=0.42]{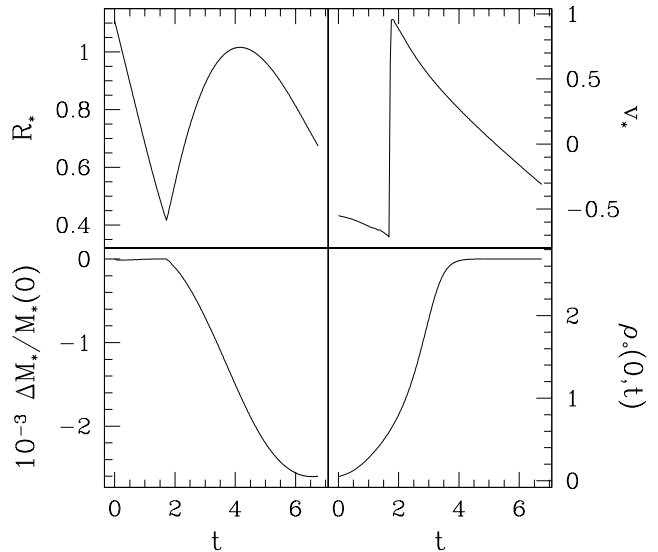}
\caption{Evolutions of stellar radius ($R_\star$), velocity 
at $R_\star$ ($v_\star$), relative stellar mass 
deviation from initial time ($\Delta M_\star(t)/M_\star(0)$), and central density for a SBC star.  
The evolution was stopped when the maximum value 
of $2m/r$ obtained a value of $0.995$, 
at which point the mass of black hole was calculated to be about $0.1080$ and the minimum of $\alpha$ 
was $1.0\times 10^{-8}$.  The defining parameters for this run are 
$\rho_\circ(0,0) = 0.05$, $v_{min}(0) = -0.556$, and $M_\star(0) = 0.1092$.  
\label{fig:sbc-bound}}
\end{figure}

A dynamical scenario is said to be of the type SBC if a black hole forms, a shock/bounce occurs, 
and $\Delta M_\star(t) / M_\star(0)$ decreases over the entire course of the evolution by 
an amount greater than $10$ times the numerical error in calculating 
$\left(\Delta M_\star(t) / M_\star(0)\right)$.  The numerical error here is the 
timestep-to-timestep stochastic fluctuation we see in this quantity due to truncation and roundoff errors.  
The distinction between SBC and PC states is somewhat arbitrary 
because we are unable to measure the eventual steady-state mass 
of a nascent black hole, due to restrictions imposed by our coordinate system.
Further, we do find a few instances where the star's matter is still 
trapped even after the shock and bounce, as seen in Fig.~\ref{fig:sbc-bound}.  
That is, the external matter bounces from the denser core, 
forms a shock and propagates outwards, but a portion of this 
matter eventually falls back onto the black hole.  
The fact that $R_\star$ decreases 
and $v_\star$ becomes ingoing after the bounce suggests that the outer parts of the star do
indeed accrete onto the collapsing interior.  
This example demonstrates that not all SBC scenarios
result in black holes that are less massive than their progenitors, and 
that the final mass of the black hole is most likely continuous across the SBC/PC boundary.

For less compact stars, it is natural to justify the existence of the transition between
SBD to SBO scenarios.  If we follow evolutions of a particular star---say one with $\rho_c=0.03$---
for various $v_{\min}$, we see that the initial velocity perturbation results in dispersal of 
more and more of the stellar material as $v_{\min}$ increases.  The central 
densities and masses of the resultant SBO stars decrease as the SBO/SBD boundary is reached, 
implying that the transition is continuous.  For instance, if $\rho_c^f$ and $M_\star^f$ are
the final central density and mass, respectively, of the product star, then we should see that 
$\rho_c^f,M_\star^f \rightarrow 0$ as 
$v_{\min} \rightarrow v_{\min}^{\star-}(\rho_c)$, where $v_{\min}^\star(\rho_c)$ is the 
threshold value of $v_{\min}$ that separates the SBO and SBD states.
We have found that this seems to be the case since after
tuning $v_{\min} \rightarrow v_{\min}^\star(0.03)$ to an 
approximate precision of $10\%$, $\rho_c^f \simeq 0.0045$---which is about an $85\%$ decrease
in central density.  Alternatively, we cross the threshold by varying $\rho_c$ 
and keeping $v_{\min}$ constant.  That is, if we choose a 
specific $v_{\min}$ and start perturbing stars with larger $\rho_c$,
we see that---as
the stars become less compact---the velocity distribution is able to expel more and more
matter from the central core.  In turn, smaller and smaller stars will form for a given $v_{\min}$
as $\rho_c \rightarrow \rho_c^{\star+}(v_{\min})$, where $\rho_c^{\star}(v_{\min})$ is the value of $\rho_c$ at
the SBO/SBD boundary for a given value of $v_{\min}$.  It would be interesting to calculate
the scaling behavior of $M_\star^f$ as a function of $\rho_c - \rho_c^\star(v_{\min})$ 
or $v_{\min}^\star(\rho_c)-v_{\min}$.   An accurate calculation of this
scaling law would require many runs in this regime, which is 
one of the most computational intensive regimes.  
In this limit, we would have to resolve
a wide range of scales in order to evolve the initial dynamics of the compact progenitor 
star through to it settling into a new equilibrium.  Such calculations might require a
full-fledged adaptive mesh refinement (AMR) code, which we leave for future work.

%%%%%%%%%%%%%%%%%%%%%%%%%%%%%%%%%%%%%%%%%%%%%%%%%%%%%%%%%%%%%%%%%%%%%%%%%%%%%%%%%%%%%%%
%%%%%%%%%%%%%%%%%%%%%%%%%%%%%%%%%%%%%%%%%%%%%%%%%%%%%%%%%%%%%%%%%%%%
%%%%%%%%%%%%%%%%%%%%%%%%%%%%%%%%%%%%%%%%%%%%%%%%%%%%%%%%%%%%%%%%%%%%
%%  APPENDIX D  %%%%%%%%%%%%%%%%%%%%%%%%%%%%%%%%%%%%%%%%%%%%%%%%%%%%%%%%%%
%%%%%%%%%%%%%%%%%%%%%%%%%%%%%%%%%%%%%%%%%%%%%%%%%%%%%%%%%%%%%%%%%%%%
\section{Departures of Near Critical Solutions from Unstable Equilibrium}
\label{sec:plateaus}

\begin{figure}[htb]
\includegraphics[scale=0.4]{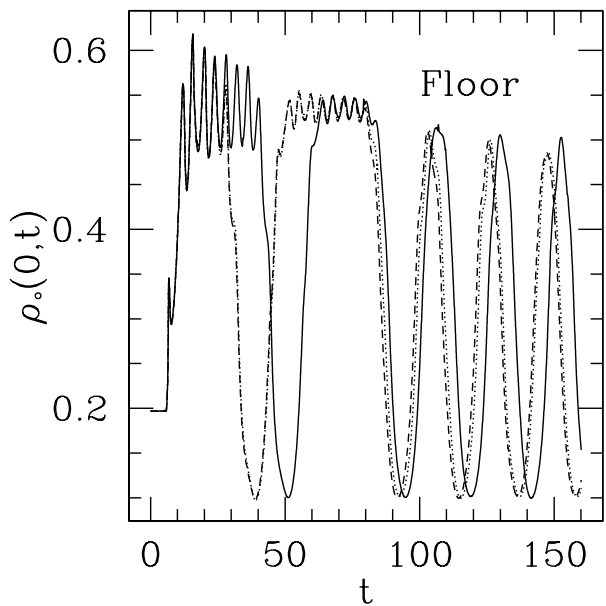}
\caption{Comparisons of $\rho_\circ(0,t)$ for 
the marginally subcritical solutions 
obtained when using varying values of the fluid's floor.  The original, reference solution (solid curve) used 
$\delta=3.8809\times10^{-18}$, while the other two lines used floor values $10$ (dotted line) and 
and $100$ (dashed line) times greater.  Variations can be seen between 
the three solutions, even though the smallest discrepancies are between the two solutions with 
the largest floor values.  All runs shown here used $\rho_\circ(0,0) = 0.197$.
\label{fig:rhoc-floor}}
\end{figure}

\begin{figure}[htb]
\includegraphics[scale=0.4]{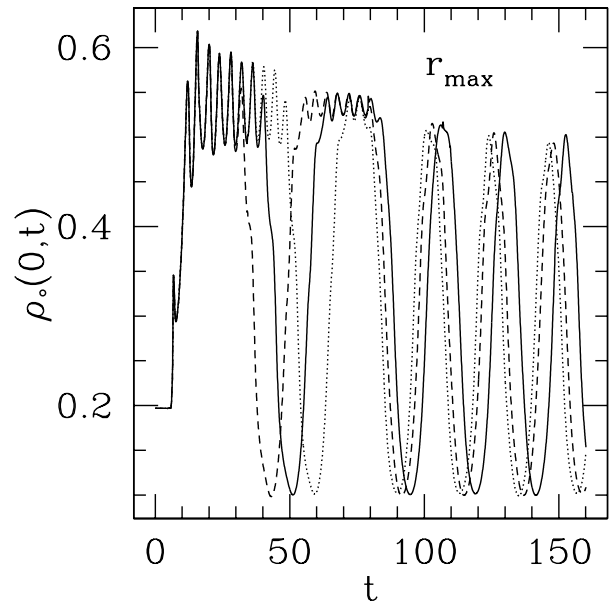}
\caption{Central density as a function of time of the subcritical solutions closest to the threshold
obtained with physical domains of various sizes.  The 
dotted (dashed) curve used a domain twice (thrice) as large as that of the original configuration,
which is shown here as a solid curve.  All runs shown here used $\rho_c = 0.197$.
\label{fig:rhoc-rmax}}
\end{figure}

\begin{figure}[htb]
\includegraphics[scale=0.4]{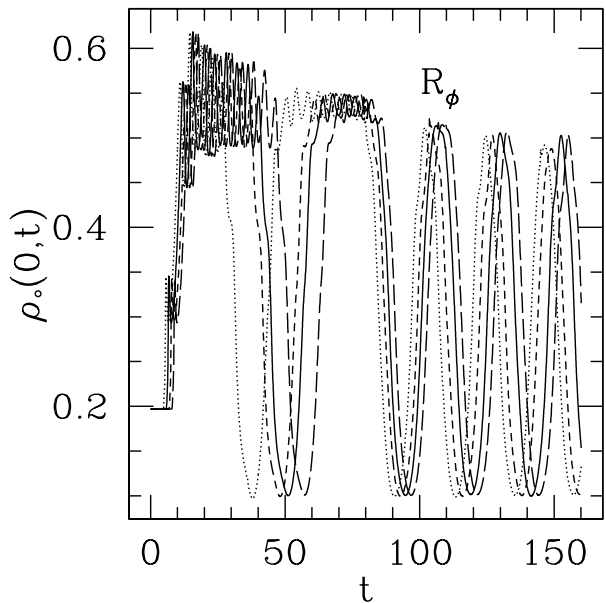}
\caption{Central 
density as a function of time of the subcritical solutions closest to the threshold
obtained by using different initial locations of the initial scalar field distribution, $R_\phi$. 
Specifically, the scalar field at $t=0$ takes the form of a Gaussian distribution, and the 
position of the center of this Gaussian is unique for each curve shown here.  In our units 
the radius of the progenitor star was $r=0.87$, while the initial positions 
of the scalar field pulses were at $r=4$ (dots), $r=5$ (solid curve), $r=6$ (short dashes), 
$r=7$ (long dashes).   All runs shown here used $\rho_c = 0.197$.
\label{fig:rhoc-r0pp}}
\end{figure}

In order to gather a better understanding of what causes the near critical solutions to temporarily 
depart from the unstable branch, we tuned to the critical solution using different 
values of various control parameters.
For instance, to see if the presence of the departures is affected by the 
floor, we tuned to the critical solution for three different values of
$\delta$.  The most marginally subcritical solutions from these 
searches are shown in Fig.~\ref{fig:rhoc-floor}.  In addition, the effect of changing the outer boundary's
location, $r_{\max}$ is seen in Fig.~\ref{fig:rhoc-rmax}.   To see if the 
time at which the pulse collides with the star has any effect, the initial position of the pulse, 
$R_\phi$ was varied; the results from this particular analysis are shown in Fig.~\ref{fig:rhoc-r0pp}. 

In general, we see all these aspects to have significant and non-trivial effect on the 
threshold solution's departure from the unstable solution. But, all the different marginally-subcritical 
solutions finally depart from the unstable solution at approximately the same time and all cases 
share the same scaling exponent.

Whether because of its magnitude or extent, 
the solution's departure seems to be affected by the floor.  Increasing the size of the floor seems
to hasten the initial departure; even though they represent only two points of reference, 
the similarity of the solutions with the two highest floor values may suggest that the floor's effect
``converges'' to one behavior as its size increases. 
On the other hand, changes in the size of the computational domain and $R_\phi$ 
seem to have no \emph{consistent} effect on the first departure time.  

The most likely explanation is that excited modes from the 
the artificial atmosphere surrounding the star instigate the departures.  
The unstable solutions to which the near critical 
solutions emulate are $1$-mode unstable TOV solutions, and TOV solutions do not involve an
atmosphere.  Since these additional modes have no effect on the scaling 
exponent and only periodically affect the evolution of the threshold solutions, they must be transient
and stable. Their little influence is consistent with the idea that they come from the atmosphere
since it is hydrodynamically and gravitationally insignificant compared to the star.  
Further work will need to be done in order to definitively understand the cause of the departures. 

Similar studies 
(e.g., \cite{jin-suen-2006,wan-jin-suen-2008,liebling-lehner-neilsen-palenzuela-2010,wan-submitted-2010,wan-thesis-2010,radice-2010,kellerman-2010}) 
have not reported encountering similar phenomena.  It remains to be seen whether it is because their investigations 
involved different neutron star solutions (e.g., different values of $K$ as in \cite{jin-suen-2006}, or rotating and magnetized
unstable-branch neutron stars as in \cite{liebling-lehner-neilsen-palenzuela-2010}), different numerical schemes, or something else altogether. 

%%%%%%%%%%%%%%%%%%%%%%%%%%%%%%%%%%%%%%%%%%%%%%%%%%%%%%%%%%%%%%%%%%%%%%%%%%%%%%%%%%%%%
%%%%%%%%%%%%%%%%%%%%%%%%%%%%%%%%%%%%%%%%%%%%%%%%%%%%%%%%%%%%%%%%%%%%%%%%%%%%%%%%%%%%%
%%%%%%%%%%%%%%%%%%%%%%%%%%%%%%%%%%%%%%%%%%%%%%%%%%%%%%%%%%%%%%%%%%%%%%%%%%%%%%%%%%%%%
%%%%%%%%%%%%%%%%%%%%%%%%%%%%%%%%%%%%%%%%%%%%%%%%%%%%%%%%%%%%%%%%%%%%%%%%%%%%%%%%%%%%%
%%%%%%%%%%%%%%%%%%%%%%%%%%%%%%%%%%%%%%%%%%%%%%%%%%%%%%%%%%%%%%%%%%%%%%%%%%%%%%%%%%%%%
%%%%%%%%%%%%%%%%%%%%%%%%%%%%%%%%%%%%%%%%%%%%%%%%%%%%%%%%%%%%%%%%%%%%%%%%%%%%%%%%%%%%%
%% BIBLIOGRAPHY:   %%%%%%%%%%%%%%%%%%%%%%%%%%%%%%%%%%%%%%%%%%%%%%%%%%%%%%%%%%%%%%%
%%%%%%%%%%%%%%%%%%%%%%%%%%%%%%%%%%%%%%%%%%%%%%%%%%%%%%%%%%%%%%%%%%%%%%%%%%%%%%%%%%%%%
%%%%%%%%%%%%%%%%%%%%%%%%%%%%%%%%%%%%%%%%%%%%%%%%%%%%%%%%%%%%%%%%%%%%%%

\end{document}